\documentclass[twocolumn,preprintnumbers,superscriptaddress]{revtex4-2}
\usepackage{hyperref}
\usepackage[utf8]{inputenc}
\usepackage{amsmath}

\usepackage{soul}

\usepackage{url}

\usepackage{listings}
\usepackage{xcolor}
\usepackage{natbib}
\usepackage{graphicx}
\usepackage{physics}
\usepackage{tikz}
\usetikzlibrary{quantikz}

\newcommand{\DSD}{
Research Group on Data Science for the Digital Society
La Salle - Universitat Ramon Llull
Carrer de Sant Joan de La Salle, 42
08022 Barcelona (Spain)
}
\newcommand{\LDIG}{
Lighthouse Disruptive Innovation Group, LLC
7 Broadway Terrace, Apt 1
Cambridge MA 02139
Middlesex County, Massachusetts (USA)
}
\newcommand{\UVA}{
Universidad de Valladolid
C/Plaza de Santa Cruz, 8, 
47002 Valladolid (Spain)
}

\newcommand{\orcid}[1]{\href{https://orcid.org/#1}{\includegraphics[width=8pt]{orcid.png}}}

\graphicspath{{./images/}}

\begin{document}

\title{qRobot: A Quantum computing approach in mobile robot order picking and batching problem solver optimization}
\author{Parfait Atchade-Adelomou}
\affiliation{\DSD}
\email{parfait.atchade@salle.url.edu}
\affiliation{\LDIG}
\email{parfait.atchade@lighthouse-dig.com}

\author{Guillermo Alonso-Linaje}
\affiliation{\UVA}
\email{guillermo.alonso.alonso-linaje@alumnos.uva.es}

\author{Jordi Albo-Canals}
\affiliation{\LDIG}
\email{jordi.albo@lighthouse-dig.com}

\author{Daniel Casado-Fauli}
\affiliation{\DSD}
\email{daniel.casado@salle.url.edu}

\date{May 2021}

\begin{abstract}
This article aims to bring quantum computing to robotics. A quantum algorithm is developed to minimize the distance travelled in warehouses and distribution centres where order picking is applied. For this, a proof of concept is proposed through a Raspberry Pi 4, generating a quantum combinatorial optimization algorithm that saves the distance travelled and the batch of orders to be made. In case of computational need, the robot will be able to parallelize part of the operations in hybrid computing (quantum + classical), accessing CPUs and QPUs distributed in a public or private cloud. Before this, we must develop a stable environment (ARM64) inside the robot (Raspberry) to run gradient operations and other quantum algorithms on IBMQ, Amazon Braket, Dwave and Pennylane locally or remotely. The proof of concept will run in such quantum environments above. 
\newline
\newline
\textbf{KeyWords:} Quantum Computing, Machine Learning, Picking Problem, Batching Problem, Quantum Robotics, Raspberry PI4, Docplex
\end{abstract}

\maketitle

\section{Introduction}\label{sec:introduction}
From DHL, Gartner and others \cite {angeleanu2015new, kuzmicz2015benchmarking, savelsbergh201650th}, we know that the first wave of automation using smart robotics has reached the logistics industry. Driven by rapid technological advancements and increased affordability, robotic solutions (software and hardware) are forcibly entering labour logistics, supporting flawless processes and boosting productivity. Robots, especially mobile, will adopt more roles in the supply chain, helping workers with storage, transportation and little by little, they will expand their service. In fact, in some countries, there are already robotic delivery services \cite{while2021urban}.

We are already living an exponential increment of mail-order shopping, online shopping and supply chain systems, requiring large-scale logistic centers. Almost everyone can order products remotely, and the logistic center increases its functionalities, including keeping and shipping products. 
While there was a tendency to increase the adoption of automated systems based on robots powered by AI to increase efficiency \cite {van2018robotic, siderska2020robotic, agostinelli2020towards}, COVID-19 introduced the concept of touch-less online shopping that reduces the risk of infections. Smart Warehouses are the epicenter of the cost-efficiency of any e-commerce company \cite{Tompkins2010planning}. 

The emerging field of hybrid (quantum-classical) algorithms joins CPU, and QPU \cite {Karalekas2020} to speed up specific calculations within a classical algorithm. This allows for shorter quantum runs that are less susceptible to the cumulative effects of noise and work well in current devices.
This article is intended to explore the performance of a quantum picking model. A hybrid system is proposed that effectively replaces the current ones and opens the doors to quantum computing in robotics.

After Section \ref{sec:introduction}, the document is organized as follows; Section \ref{sec:Related_work} shows previous work on both assembly techniques and approaches to picking and batch management systems; then, Section \ref{sec:QCircuit_NISQ} presents the quantum fundamentals needed from this era to solve this problem; next, the implementation of the proposed strategy and the creation of the qRobot performed in Section \ref{sec:implementation} are explained; to continue, Section \ref{sec:result}, which shows the results of our experimental analysis, and Section \ref{sec:Discussions}, in which some open problems are summarized, compared and presented; and finally, Section \ref{sec:Conclusions} concludes the previous results and describes the future work.

\section{Work Context}\label{sec:Related_work}
According to \cite{chen2016cancer,bustillo2015slaughterhouse,koch2016grouping}, supply chains, warehouses and distribution centres occupy a very important position when storing and serving customer demand. Today, in order to be competitive within this sector, Logistics 4.0 has been created, which is known as the set of artificial intelligence technologies and techniques that seek the efficiency of the movements of materials and products in a factory or warehouse. Better time management helps logistics companies find and locate a material, reduce fatigue and possible workplace accidents, and spend less time documenting items.

Many works of literature highlight these factors as the main ones where the loss of time and resources in a process require an urgent solution, and precisely, it is technologies such as Artificial Intelligence and the Internet of Things (IoT), which today allow us to optimize them\cite{albareda2009multi,cergibozan2019order,azadnia2013order}.

Only in the last decade, researchers focused on addressing the multiple order picking planning problems. The study of the efficiency of a Warehouse can be addressed based on multiple parameters. According to \cite{vangils2018picking}, there are three key considerations: 1) Performance Measure (time, cost, productivity, and service), 2) How we model the warehouse (Analytical model, Mathematical Model, or Simulation), and the combination of factors (storage location assignment, routing, order batching, or other order picking planning problems).

Based on data from \cite{vangils2018picking}, we can see the percentage of relevance of the considered order picking planning problems based on the percentage of papers that are related to such challenges: 

\begin{figure}[!ht]
\centering
\includegraphics[width=0.45\textwidth]{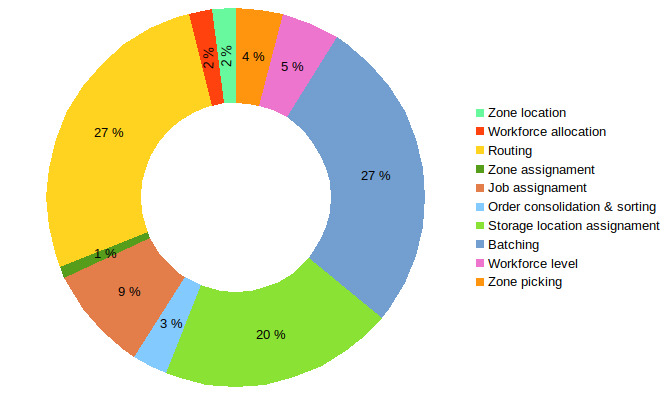}
\caption{Distribution of considered order picking problems based on percentage of publications.
}
\label{fig:planningprob}
\end{figure}

As we can see in Fig.\ref{fig:planningprob} Picking and Batching are the top priorities based on the research contributions. 

Order preparation (picking) is one of the most frequent and costly operations in labour \cite{chen2016cancer,  bustillo2015slaughterhouse} since it is responsible for recovering the items required by the orders of customer orders (could also be supplied, but in this article, we focus exclusively on sales orders). And to create the batches, grouping several orders of orders in a picking list to collect all the batch demands in a single warehouse tour. In this last part of order preparation, our quantum algorithm comes into action to optimize the routes travelled to achieve efficient picking.

There are many techniques and strategies for solving the picking problem. The most striking are “The selected techniques for evaluation include A *”\cite{duchovn2014path}, “Potential Fields (PF)”, “Rapidly-Exploring Random Trees * (RRT *)”\cite{lavalle2001rapidly, lavalle1998rapidly, cheng2002resolution} and "Variations of the Fast-Marching Method (FMM)"\cite {rawlinson2005fast}.
Other strategies have explored using the TSP and the VRP as algorithms to solve the picking problem. In this case, if the number of order orders per lot is greater than two \cite {gademannvan}, picking becomes an NP-Hard problem in which the number of possible lots and binary variables increase exponentially with the number of purchase orders \cite {gademannvan}.
From there, several heuristic techniques, methods and algorithms (for example, genetic) were born to relax these difficulties \cite{Cortina2001,azadnia2013order,chen2016cancer,hsu2005batching,koch2016grouping,tsai2008using}. However, and as mentioned above, depending on the volume of data, the computational cost of the algorithm becomes intractable for classical computing.

The latter leads us to explore new approaches to the large-scale picking problem, and one of the approaches to take into account to solve this task is quantum computing [6].
Quantum computing could help us change the degree of complexity of the problem, enhanced by its high computing power. Among the great fields where quantum computing is called to stand out is constraint satisfaction problems (CSP) \cite{tsang2014foundations}. One of the useful algorithms in this field is Quadratic Unconstrained Binary Optimization (QUBO) problems \cite{kochenberger2014unconstrained}.

From Alan Turing, \cite {turing1937computable} to Richard Feyman's idea of considering the simulation of systems in quantum mechanics by other quantum systems \cite {feynman1982simulating}, interest in creating new ways of solving them has grown dramatically. This, together with the consequences of the well-known Moore's Law, gave way to the idea of building quantum computers. Over the past decades, before demonstrating the superiority of quantum computing, David Deutsch published this article\cite {deutsch1985quantum} in which he proposed how a universal quantum computer could be. Years later, the worth of these new computers has been demonstrated to solve some specific problems such as factoring prime numbers using Shor's \cite {shor1994algorithms} algorithm or searching in disordered sets with Grover's \cite {grover1996fast} algorithm, although all this limited to the number of qubits available. We are currently in the NISQ era \cite {Joh18} in which we have computers between 50 and 100 qubits (Gate-Based Quantum Computer), opening the way to the emerging field of hybrid quantum-classical computing. Within this, different algorithms have been developed, such as “VQE” \cite {Dao19}, “QAOA” \cite {farhi2014quantum} or, “Quantum Machine Learning (QML)” \cite {Mar14, JBi17, Adr20, adelomou2020using, atchadeadelomou2021quantum}, which we will focus on with this article.

There are two dominant techniques for quantum computing. Continuous-Time Quantum Computing \cite{Kendon2020} used by D-Wave in which the problem to solve is mapped in quantum hamiltonians and the natural dynamics of physical systems, and the Gate based Quantum Computing\cite{McG14, Mic00, Kir17} led by IBM, in which the computation is made through a series of discrete gate operations. The Ref.\cite{Kendon2020} argues how Quantum Walk (QW), Quantum Annealing (QA), and Adiabatic Quantum Computing (AQC) are related. QW and AQC are pure quantum evolutions (unitary), while QA involves external cooling. 

The Adiabatic Quantum Computing proposed by Farhi \cite{Edw,Edw19}, is based on the adiabatic theorem \cite{McG14} and was the first quantum computing technique. 

Quantum Annealing, based on the adiabatic quantum computing paradigm, was initially introduced by Kadowaki and Nishimori \cite{Nis08}. Since its proposal, the QA technique was a light for solving combinatorial optimisation problems. This technique tries to solve problems similar to how optimisation problems are solved using the classical simulated annealing \cite{McG14}. From a multivariate function formed from an energy landscape so, the ground state corresponds to the optimal solution of the problem. The QA process must be repeated until finding the optimal solution to the problem. The most significant advantage of quantum annealing is its high degree of parallelism over classical code execution. Because it analyses all possible inputs in parallel to find the optimal solution, this is very useful when we want to reduce the complexity of the NP-complete problems.

QA has confirmed its ability to solve a broad range of combinatorial optimisation problems. And also in other fields, such as quantum chemistry (One of the fields that are taking great advantage of capacity and the era in which quantum computing is right now) \cite{McG14}, bioinformatics\cite{McG14} and routing \cite{MarPs2}, to cite a few.

We can categorise combinatorial optimisation problems into several groups. Where the need for adequate techniques for solving such problems. One of the standardised optimisation problems is the aforementioned QUBO\cite{McG14, Nis08,KBe19}.

QUBO, as NP-hard, refers to a pattern matching technique that, among other applications, can be used in machine learning and optimisation and which involves minimising a quadratic polynomial on binary variables\cite{McG14}. QUBO has demonstrated its potential in solving some standard combinatorial optimisation problems such as the colouring of graphics, workshop planning, vehicle routing and programming, neural networks, the partition problem, 3-SAT, and machine learning where the parameters of the problem can be expressed as Boolean variables \cite{McG14,Nis08,KBe19}. Only to remember that Adiabatic quantum annealing techniques are also used to solve multi-objective optimisation problems\cite{BOm04}. The QUBO formulation is suitable for running a D-Wave architecture; nevertheless, QUBO can be mapped on the Ising model\cite{McG14}.

Advances in quantum computing offer a way forward for efficient solutions to many cases of substantial eigenvalue problems unsolvable in a traditional way \cite{Alb13}. Quantum approaches to finding eigenvalues previously relied on the Quantum Phase Estimation (QPE) algorithm. The QPE is one of the essential subroutines in quantum computation. It serves as a central building block for many quantum algorithms and offers exponential acceleration compared to classical methods, and requires several quantum operations  $O \left( \frac{1}{p} \right)$  to obtain an estimate with precision  $p$  \cite{Alb13,GGG19}. 

Variational Quantum Eigensolver (VQE) proposed by Peruzzo\cite{Alb13} based on the variational principle and form, estimates the ground state energy of the Hamiltonian of the problem \cite{Jer03}. The VQE is a hybrid quantum/classical algorithm originally proposed to approximate the ground state of a quantum system (the state attaining the minimum energy).
Quantum Approximate optimisation Algorithms (QAOA), based on the principles of adiabatic quantum computation \cite{McG14, GGG19, Qin18}, is used to solve QUBO problems. Farhi and Harrow showed the advantages of QAOA compared to classical approaches \cite{Edw, Edw19}. While Rebentrost \cite{Pat19} just debated the problems of constrained polynomial optimisation using adiabatic quantum computation methods. Other scientists such as Vyskocil and Djidjev \cite{Tom19} worked on how to apply restrictions in QUBO systems to avoid the use of large numbers of the coefficients so, thus more qubits, resulting from the use of quadratic penalties, they proposed a new combinatorial design which involved solving problems of linear programming of mixed integers to adapt applications restitution. Anuradha Mahasinghe, Richard Hua, Michael J. Dinneen, and Rajni Goyal\cite{Anu19} investigated and solved the Hamiltonian cycle problem in computational frameworks such as quantum circuits, quantum walks, and adiabatic quantum computing. 
All of these advances in quantum computing have been applied to routing and scheduling techniques. The researcher Lucas contributed an expansive vision and discussions on Ising formulations for various NP-complete and NP-hard optimisation problems, emphasising using as few as possible qubits. In the same way, there have been many works of literature on the VRP \cite{Seb19} and its variants.

\begin{figure}[!ht]
\centering
\includegraphics[width=0.45\textwidth]{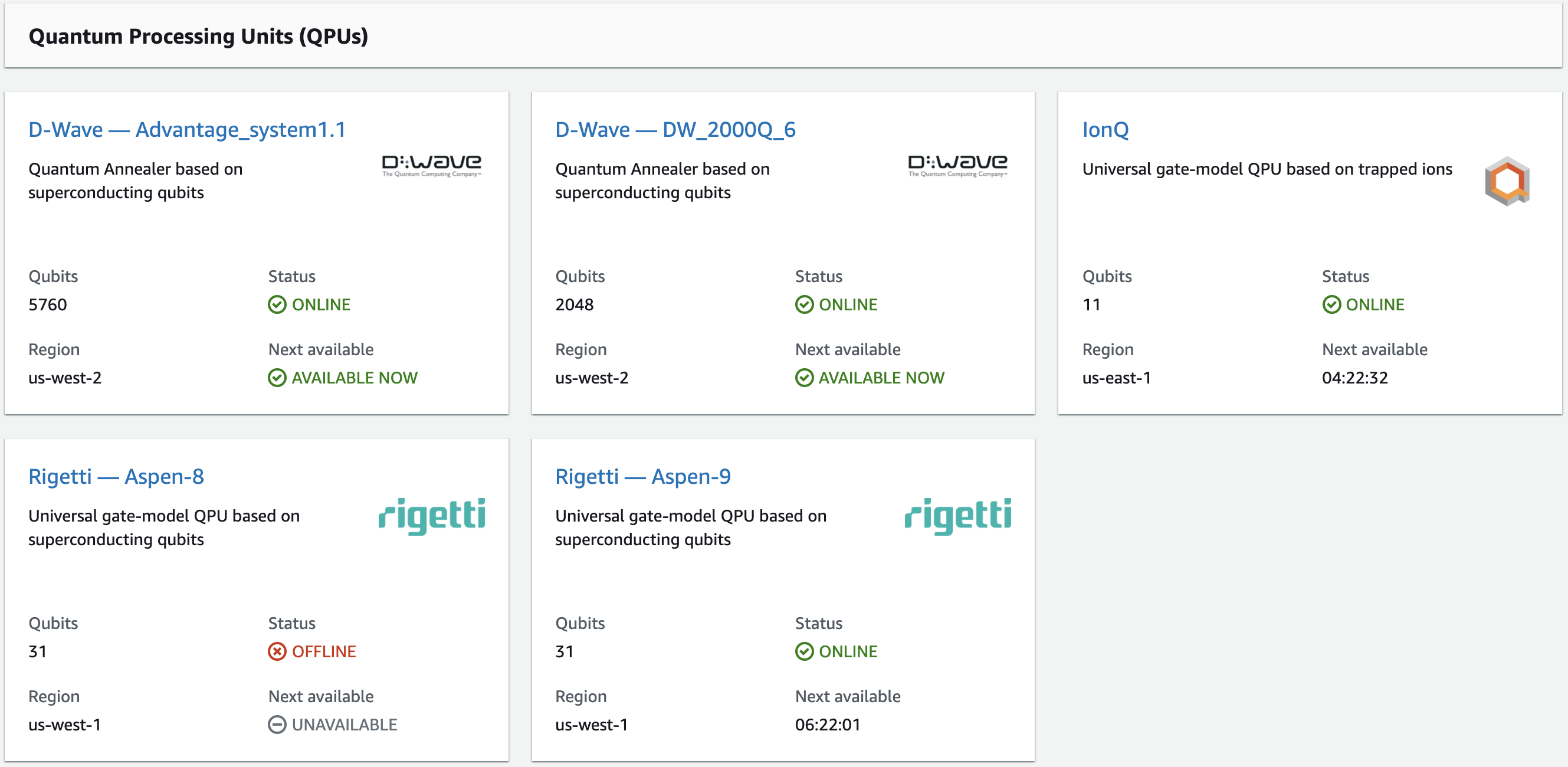}
\caption{The Quantum Hardware Technologies (Gate-based superconducting qubits, Gate-based ion traps and Quantum annealing) available in Amazon Braket.}
\label{fig:AWS_QC}
\end{figure}

\begin{figure}[!ht]
\centering
\includegraphics[width=0.45\textwidth]{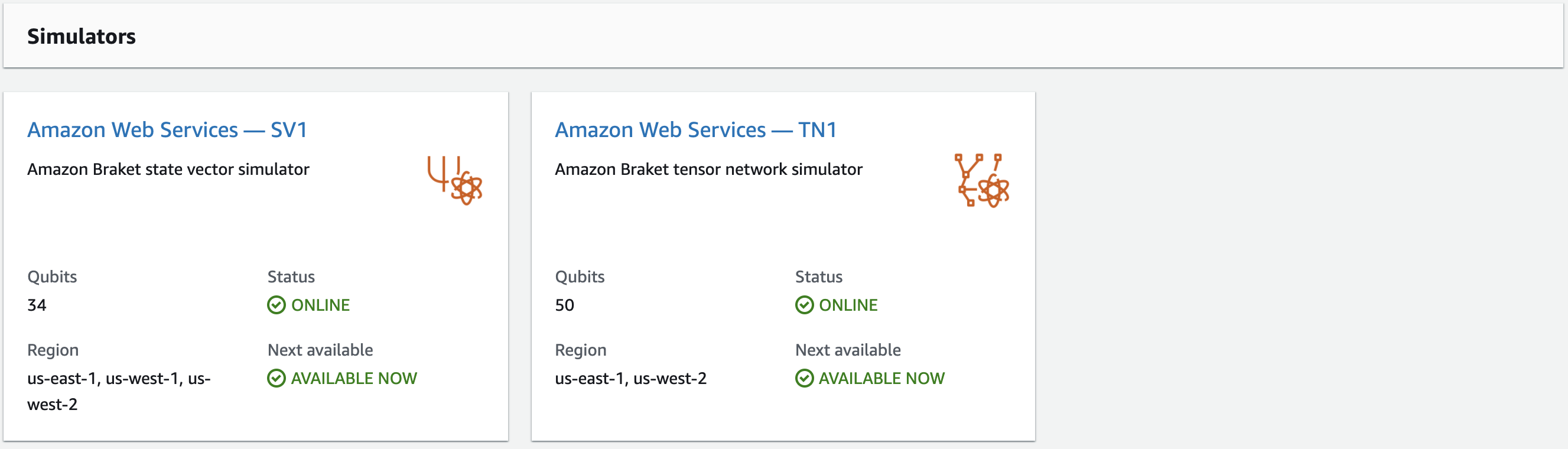}
\caption{Quantum Simulators systems where we can see it state vector simulator (34 qubits) and tensor network simulator (50 qubits).}
\label{fig:AWS_Simulador}
\end{figure}

Amazon Braket\cite{AWS_Braket_web} is a cloud-based (Fig.\eqref{fig:AWS_QC} and Fig.\eqref{fig:AWS_Simulador}), fully managed quantum computing service that helps researchers and developers get started into quantum world technology to accelerate research and discovery. Amazon Braket provides a development environment to explore and create, test and run quantum algorithms, quantum circuit simulators, and different quantum hardware technologies. 

We will take advantage of all these related works to define an appropriate strategy for our proposal in this NISQ era.

Studying and comparing different optimisation methods of warehouse's challenge, like picking and batching, \cite{vangils2018picking} propose three options: analytical models, simulation experiments, and mathematical programming. In our approach, we consider the latter. We use a set of mathematical expressions that describe the problem, represented by an objective mathematical function and constraints within the classical context and translate it to the quantum domain.  

While reviewing state of the art, this reference \cite {xie2021formulating} was found. The integrated order routing and the batch problem is modelled in such systems as an extended multi-tank vehicle routing problem with network flow formulations of three indices and two commodities. Such a variable neighbourhood search algorithm provides close to optimal solutions within a computational time acceptable for classical but not quantum computing.

This article intends to bring quantum computing to robotics by proposing an approach that combines the experience of classical robotics computing with the computation of complex and high-cost processes by quantum computing. We suggest preparing an environment to execute the quantum algorithms in the mobile and autonomous robot remotely and locally and design a quantum algorithm that helps the efficiency of the warehouse management.

\section{Quantum Circuits in the NISQ era}\label{sec:QCircuit_NISQ}
Quantum circuits are defined mathematically as actions in an initial quantum state. Quantum computing largely uses quantum states constructed from qubits, namely, binary states represented by $ \ket {\psi} = \alpha \ket{0} + \beta \ket{1} $. Its number of qubits $n$ commonly defines the states of a quantum circuit, and normally, the initial state of the circuit $ \ket \psi_ {0} $ is the zero state $ \ket{0} $. Mostly, a quantum circuit implements an internal unit operation $U$ in the initial state $\ket \psi_ {0}$ to transform it into the final output state $ \ket \psi_{f} $ . This $U$ gate is normally fixed and is known for algorithms or problems. In contrast, others define its internal operation through a fixed structure, called Ansatz \cite {Ansatz_best}, and adjustable parameters $\theta$ \cite {Suk191}. Parameterized circuits are beneficial and have interesting properties in this quantum era, as they broadly define the definition of ML and offer the flexibility and viability of unit operations with arbitrary accuracy \cite {JBi17, Atc20, Adr20, Mar14}.

\begin{figure*}[t!]
\centering
\includegraphics[width=.7\textwidth]{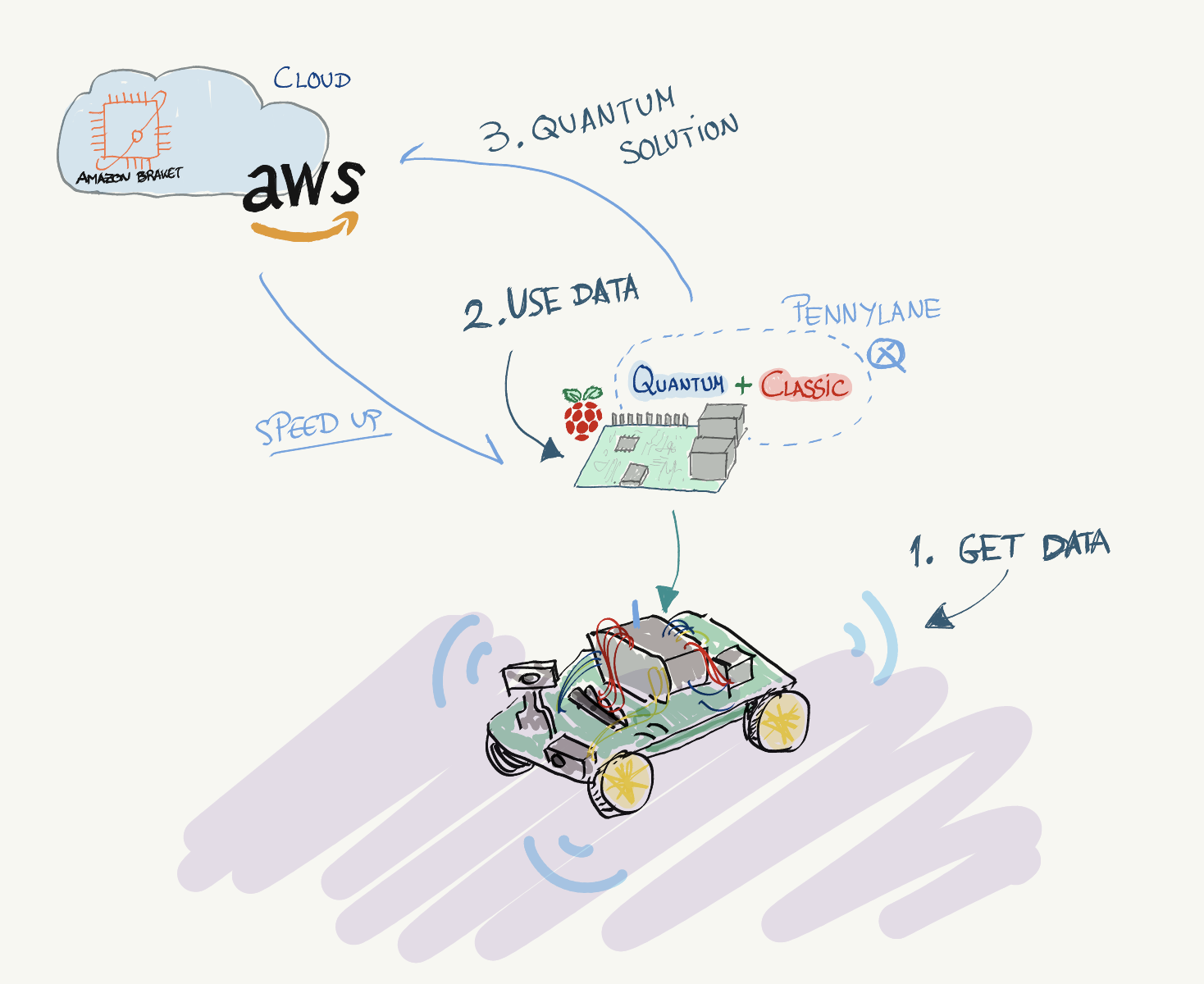}
\caption{We propose a robot that prepares batches and increases the efficiency of picking in a warehouse, taking advantage of the classic Machine Learning experience and leveraging hybrid computing (classical + quantum) in the cloud and distributed. This robot uses the Optimal routing strategy to calculate the shortest route, regardless of the layout or location of the items.}
\label{fig:qRobot}
\end{figure*}

\subsubsection{Variational Quantum Eigensolver}
\begin{figure}[!ht]
\centering
\includegraphics[width=0.45\textwidth]{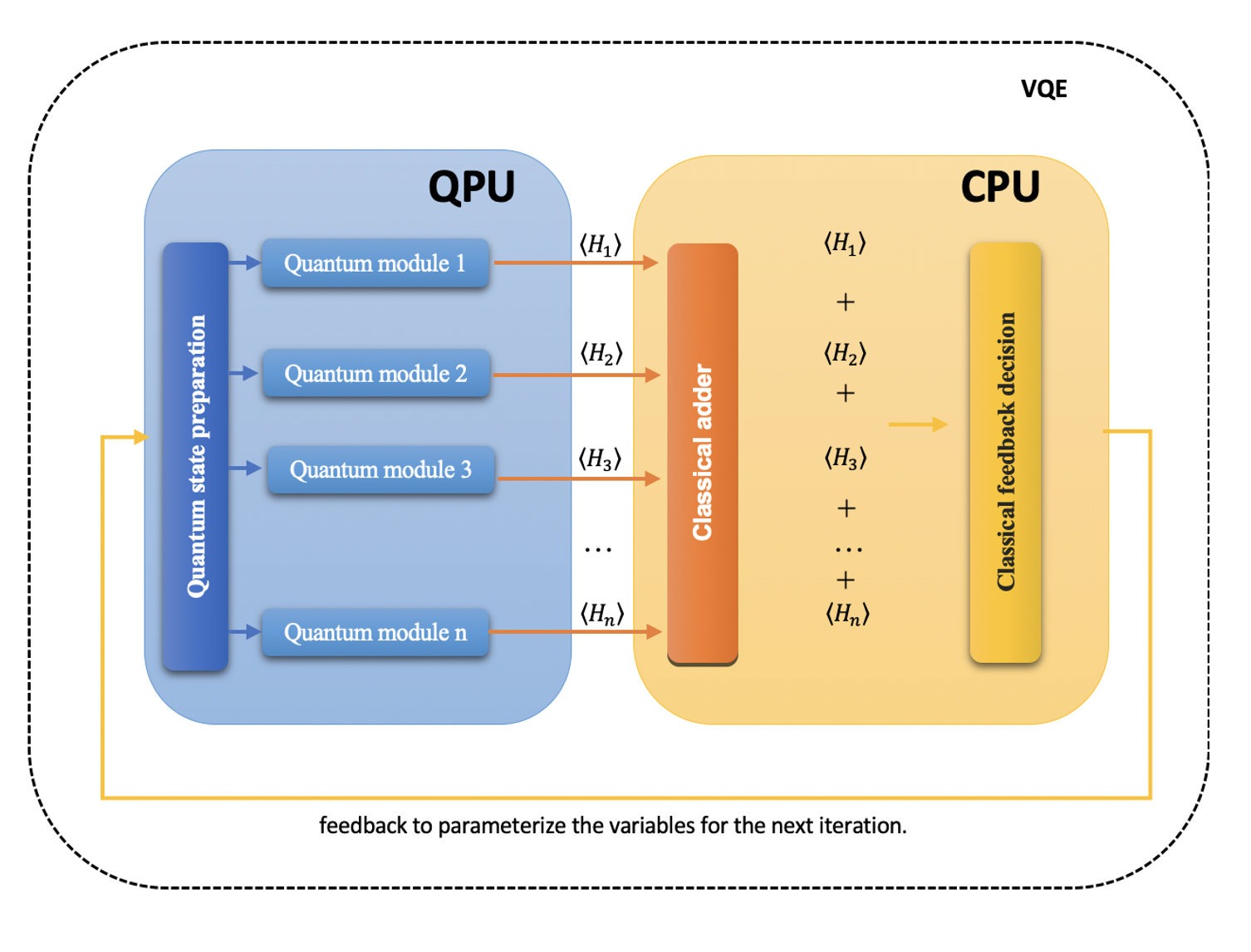}
\caption{The Variational Quantum Eigensolver diagram.}
\label{fig:VQE}
\end{figure}

\begin{figure}[!ht]
\centering
\includegraphics[width=0.45\textwidth]{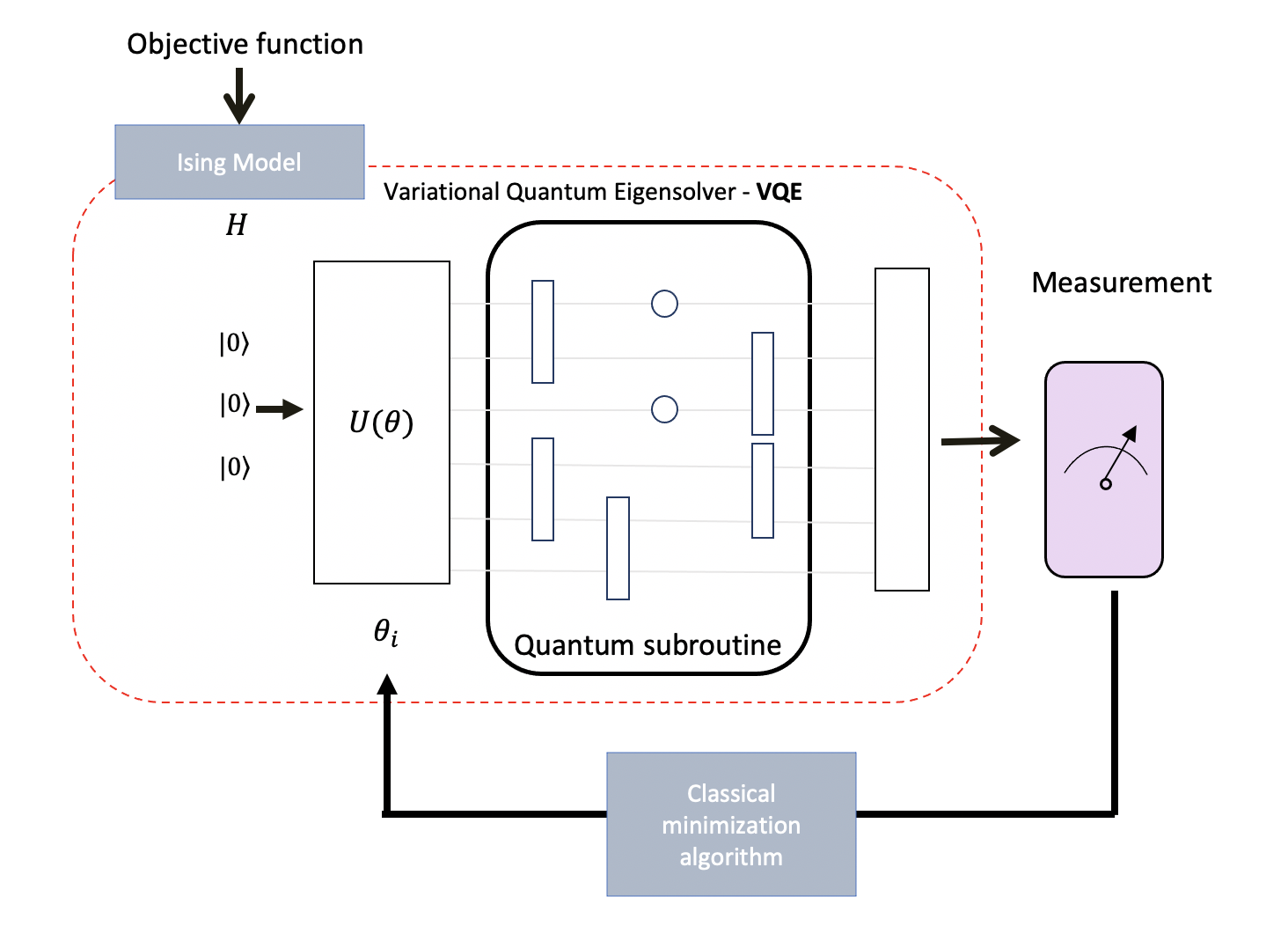}
\caption{VQE working principle based on the quantum variational circuit.}
\label{fig:VQE_f}
\end{figure}

The Variational Quantum Eigensolver (VQE) \cite{Dao19} is a classical hybrid quantum algorithm that combines aspects of quantum mechanics with the classical algorithm (Fig.\eqref{fig:VQE}). Its great contribution is to find approximate solutions to combinatorial problems. Its operation is based on mapping the combinatorial problems in a physics problem. That is, about a problem that can be formulated in terms of a Hamiltonian Ising model. Therefore, identifying the solution to the combinatorial problem is linked to finding the ground state of this physics problem. Thus, the goal is to find the ground state of this Hamiltonian.
The unknown eigenvectors are prepared by varying the experimental parameters and calculating the Rayleigh-Ritz ratio \cite{Wu2005} in a classical minimization (Fig. \eqref{fig:VQE_f}). At the end of the algorithm, the reconstruction of the eigenvector stored in the final set of experimental parameters that define the state will be done.

From the variational principle, the following equation $\langle H \rangle _{ \psi   \left( \overrightarrow{ \theta } \right) } \geq  \lambda _{i}$ can be reached out. With $\lambda _{i}$  as eigenvector and  $\langle H \rangle _{ \psi \left( \overrightarrow{ \theta } \right)}$  as the expected value. By this way, the VQE finds \eqref{expectative_value} such an optimal choice of parameters $\overrightarrow{\theta }$, that the expected value is minimized and that a lower eigenvalue is located. 
\begin{equation}
\label{expectative_value}
 \langle H \rangle =\langle\psi\left(\theta  \right)\vert H \vert\psi\left(\theta  \right)\rangle 
\end{equation}
We will use the VQE (Fig. \eqref{fig:VQE_f}) to find the minima of our objective function translated to the Ising model.

\section{Implementation}\label{sec:implementation}
To carry out the implementation of our proof of concept (Fig.\eqref{fig:qRobot} and Fig.\eqref{fig:qRobot_platform}), we must first prepare the programming environment.
Considering that the core of our robot will be the Raspberry Pi 4\cite{Raspberry}, the first thing to do is prepare it so that it can execute quantum algorithms with the guarantees required for the proposed application and especially for future operations on gradients. It is necessary to install an ARM64 operating system\cite{jaggar1997arm,jiang2020power} with all the needed packages to run all the required environments to carry out this project.
We took advantage of the work for Raspberry Pi Os Desktop (32-bit) on which the author describes how to install and run Qiskit - IBM's open-source quantum computing software framework\cite{Qis21}— on a Raspberry Pi to turn it into a quantum computing simulator and use it to access real IBM quantum computers. In our case, we do need ARM64 because we need to execute at least the TensorFlow’s version 3.2.1. The tasks to convert the Raspberry Pi 4 in our "quantum computer" are in the Appendix \ref{sec:InstallationARM64}.

After setting up the environment, we'll focus on designing and experimenting with the announced proof of concept.

\subsection{The problem's formulation}
In this formulation, we will seek to optimize the collection of the products and, later, we will make the batches.

\begin{figure}[t!]
\centering
\includegraphics[width=.45\textwidth]{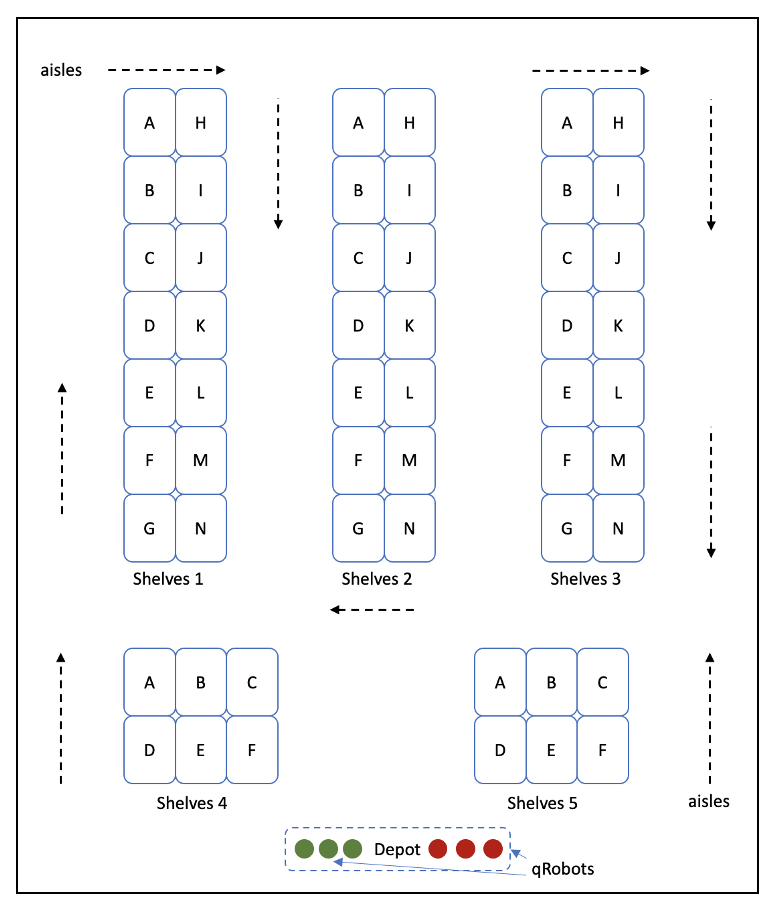}
\caption{Structure of our warehouse with pick locations. The warehouse has a rectangular layout with no unused space. We use all the parallel corridors. This proof of concept contemplates a single warehouse used to take the order and deliver it, and it is also divided into blocks, which contain slots for storing products, and transverse aisles separate them. The cross aisles do not have any products but allow the collector to navigate in the warehouse. We base our picking strategy on minimizing the route and optimizing batch preparation. We do not contemplate shelving of different levels.}
\label{fig:Wharehouse}
\end{figure}

To carry it out, we will consider the following assumptions:
\begin{enumerate}
    \item The strategy we will follow is the picking routing problem to retrieve each lot which the total distance travelled to retrieve all the items in a lot will be calculated.
    \item The warehouse configuration is given in figure\eqref{fig:Wharehouse}.
    \item For the orders of the storage positions, more than one picking robot can be used.
    \item Movements in height are not considered.
    \item Each product is stored in a single storage position, and only one product is stored in each storage position.
    \item Each picking route begins and ends at the Depot.
    \item The load capacity for each order will not exceed the load capacity of the picking robot.
    \item At the moment, the division of order orders is not contemplated. That is, only the batches of closed orders can be prepared.
    \item The concept testing will be done on all AWS-Braket, Pennylane, D-Wave and Qiskit environments. And we'll stick with the scenario that best benefits our proof of concept.
    \item We use the docplex\cite{docplex} to model our formulation.
\end{enumerate}

\begin{figure}[!ht]
\centering
\includegraphics[width=0.45\textwidth]{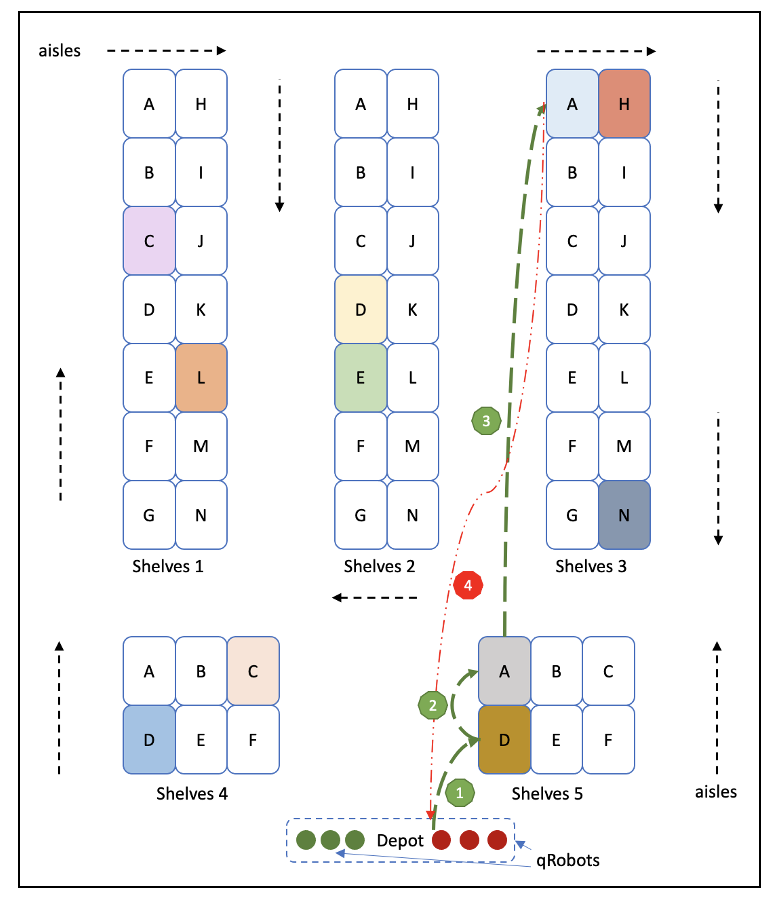}
\caption{Scenario 1, Independent lots. The robot receives the orders and calculates which order is the most optimal according to the coordinates in which each product is found. In this example, lot 4 is the most optimal.}
\label{fig:scenario1}
\end{figure}
\begin{figure}[!ht]
\centering
\includegraphics[width=0.45\textwidth]{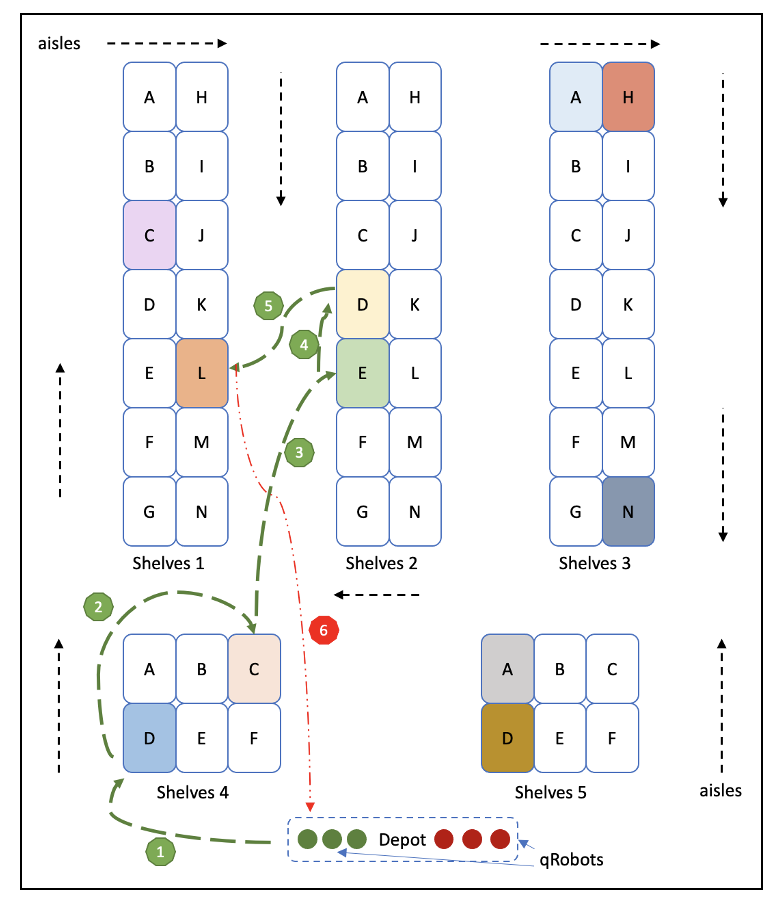}
\caption{Scenario 2, Collecting products in the same route from different batches. The robot will calculate a path that includes all the products to optimize their collection in a single journey. For example, if the product from Lot 2 is next to one from Lot 1, the robot will pick it up and store it in the basket from Lot 2.}
\label{fig:scenario2}
\end{figure}

\subsection{Picking and Batching formulation}
The formulation is represented as follows. In this scenario, the travel load is represented according to the number of robots we have. Let's imagine that we have several robots and that each of them makes a single trip. It would be the same as saying that we have a single robot that makes $n$ trips.

Let $ N_ {0} $ be the origin node, let $ N_ {1} \dots N_ {n} $ be the nodes of the products, let $ W_ {1} \dots W_ {n} $ be the weights in kg associated with For each product, let $ d_ {i, j} $ be the distance from node $ i $ to $ j $, let $ M $ be the maximum load of the qRobots, let $ K $ be the number of qRobots available, let $t$ be the instant, $i$ the node (product), and $p$ the robot and let $ x_ {t , i, p} $ our binary variable (for example, for $ x_ {2,3,2} = 1 $. It means that at time $ 2 $, the qRobot $ 2 $ is at node 3). In our formulation, time really tells us the order, that is to say $t = 0$ will be the origin $t = 1$ the moment in which it goes for the first batch. At $t = 2$ it will be the moment of the second so on.

\begin{equation}
\label{2BP_Object_Function}
\begin{aligned}
    \min_{x} \quad & \sum_{p=1}^{K}\sum_{t=1}^{n+1}{\sum_{i=0}^{n}\sum_{j=0}^{n} x_{t_{t-1},i,p}x_{t,j,p}d_{i,j}}\\
\end{aligned}
\end{equation}

\begin{equation}
\label{2BP_Object_Function_Res_1}
\begin{aligned}
\textrm{s.t.} \\
    \quad & \sum_{p=1}^{K}x_{0,0,p} = K \\
\end{aligned}
\end{equation}
\begin{equation}
\label{2BP_Object_Function_Res_2}
\begin{aligned}
    \quad & \sum_{p=1}^{K}x_{n+1,0,p} = K \\
\end{aligned}
\end{equation}
\begin{equation}
\label{2BP_Object_Function_Res_3}
\begin{aligned}
    \quad & \sum_{t=1}^{n+1}\sum_{i=1}^{n} x_{t,i,p}W{i}\leq M \qquad \forall p \in \{1, ..., K\}\\
\end{aligned}
\end{equation}
\begin{equation}
\label{2BP_Object_Function_Res_4}
\begin{aligned}
    \sum_{t=1}^{n+1}\sum_{i=1}^{n} x_{t,i,p} = 1  \qquad \forall p \in \{1,\ldots, K\}\\
\end{aligned}
\end{equation}
\begin{equation}
\label{2BP_Object_Function_Res_5}
\begin{aligned}
    \sum_{i=1}^{n} x_{t,i,p} = 1 \quad \forall t \in \{1, \ldots, n+1\} \\ \quad \forall p \in \{1, ..., K\}\\
\end{aligned}
\end{equation}

\begin{equation}
\label{2BP_Object_Function_Res_6}
\begin{aligned}
    x_{t,i,p} \in \{0,1\} \quad \forall t \in \{0, \ldots, n+1\}\\
    \\ \quad \forall i \in \{0, ..., n\}\\
    \\ \quad \forall p \in \{1, ..., K\}\\
\end{aligned}
\end{equation}

The equation \eqref{2BP_Object_Function} is our new objective function. Here we minimize the total distance. We add the distance of all the robots travelling all the time, and we will check the distances of the nodes. Restriction \eqref{2BP_Object_Function_Res_1} establishes that all the qRobots start from Depot. The restriction \eqref{2BP_Object_Function_Res_2}) establishes that all the qRobots end at Depot. The constraint \eqref{2BP_Object_Function_Res_3} establishes any robot $p$ can carry more load than allowed. The constraint \eqref{2BP_Object_Function_Res_4} declares that each robot can only be one node at any time. \eqref{2BP_Object_Function_Res_5} establish that throughout the entire route, the robots together pass each node only once and the restriction \eqref{2BP_Object_Function_Res_6} describes that $x_{t, i, p}$ are binary variables.

The number of the qubits to perform this algorithm is equal to $K(n+1)(n+2) + K \lceil log_{2}M\rceil$. At this point, we can only map our objective function in quantum and then solve it with a VQE.

\subsection {Mapping the classical to quantum optimization}
A common method for mapping classic optimization problems to quantum hardware is by coding it into the Hamiltonian\cite{eisberg1985quantum} of an Ising model \cite{lucas2014ising}.
\begin{equation}
\label{Hamiltoninano_ISing}
\begin{aligned}
    H_{Ising} = \sum_{i<j} J_{i,j}\sigma_{i}\sigma_{j} + \sum_{i} h_{i}\sigma_{i}\\
\end{aligned}
\end{equation}

Where $ \sigma_i $ is the product of $ n $ identity matrices $ I $ except a gate $ Z $ in the i-th position and $ \sigma_i \sigma_j $ product of identities minus gates $ Z $ in positions $ i $ and $ j $.

As we already can build our objective function as a QUBO in the form $\langle x^T\vert Q \vert x\rangle$, now we can map our QUBO to Ising Hamiltonian formulation leads to calculating the values of $J_{ij}$  and $h_{i}$. 

The transformation between QUBO and Ising Hamiltonian and is $z_i = 2x_i-1$, where $z_i$ is a new variable that can take the values $-1$ or $1$. This means that by writing an algorithm for QUBO with this single variable change, we will have the algorithm in Ising form. That is very useful to have the algorithm for various platforms that are based on quantum gates (IBM Q and Pennylane) or quantum annealing (meanly D-Wave) in case of going from the Hamiltonian form.
we can now solve our Picking and Batching Problem with VQE $\langle \psi(\theta)\vert H \vert \psi(\theta)\rangle$.

\section{Results}\label{sec:result}
\begin{figure*}[t!]
\centering
\includegraphics[height=4cm]{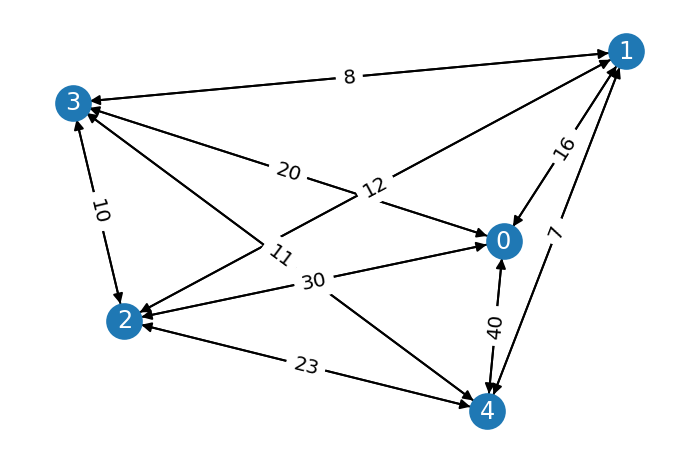}
\includegraphics[height=4cm]{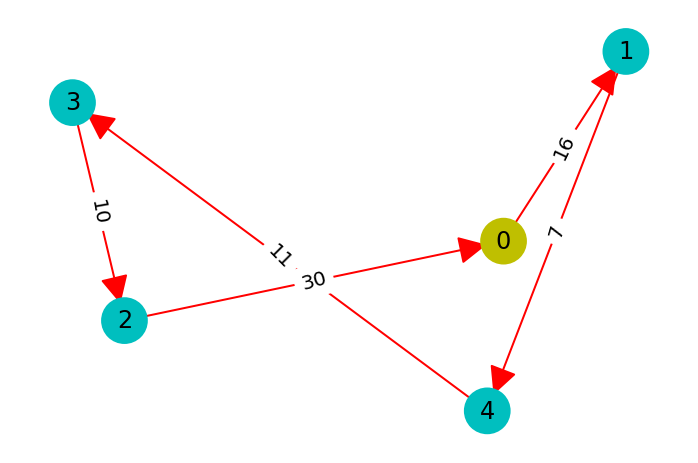}
\includegraphics[height=4cm]{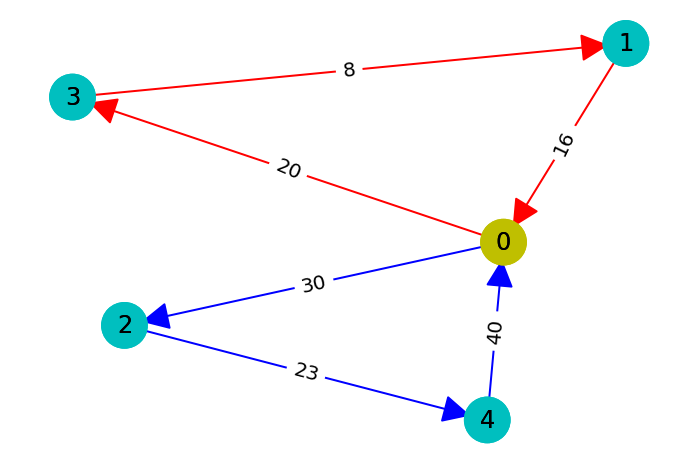}
\includegraphics[height=4cm]{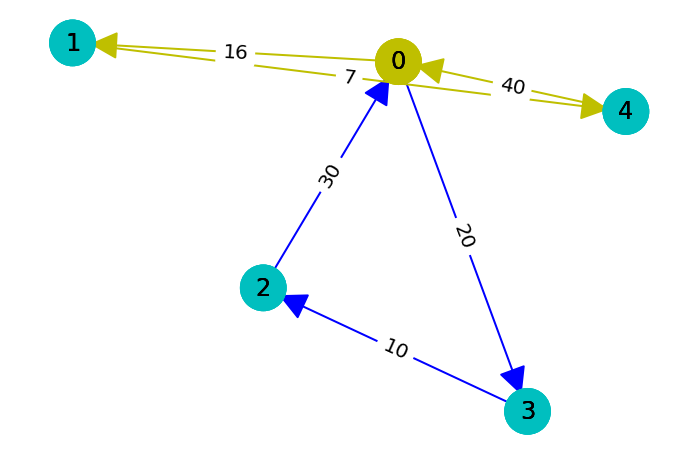}
\includegraphics[height=4cm]{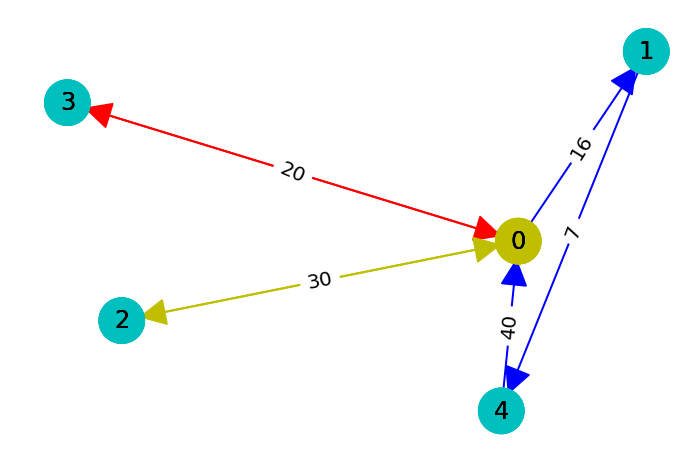}
\includegraphics[height=4cm]{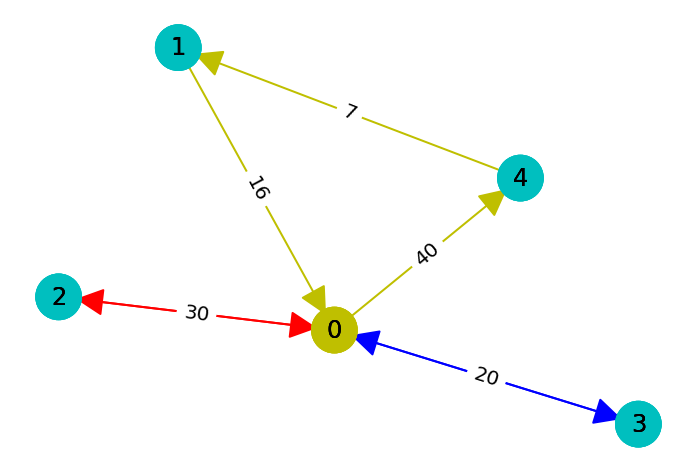}
\caption{In these graphs, we can observe the results of the algorithm in different scenarios. A different colour represents each qRobots; qRobot number $ 1 $ is red, next is blue, and the third is yellow, so on. While the depot is the $ 0 $ node in yellow color and the rest of the nodes are represented in blue. The weights of each item (not normalized) in kg are $ w_{0} = 0, w_{1} = 8, w_{2} = 8, w_{3} = 3$ and $ w_{4} = 3$. The maximum capacity of each qRobots is $ 45 $.
 In this case, we have $ 4 $ items and the possibility of using up to $ 3 $ qRobots.
Reading the images from left to right, we see that the nodes and their respective distances are shown in the first image. In the second image, show the result of the algorithm having a qRobot. In the third and fourth images, we can see two different cases solved by two qRobots. And finally, in the fifth and sixth images, we can see two other issues solved by three qRobots.
It is important to highlight that our algorithm in this proof of concept minimizes the distance travelled and optimizes the number of qRobots necessary to solve the cases presented. If it judges that the task can be performed with a single qRobot, it will not send $ 2 $ qRobots.}
\label{fig:Results_qRobots_4}
\end{figure*}

\begin{figure*}[t!]
\centering
\includegraphics[height=4cm]{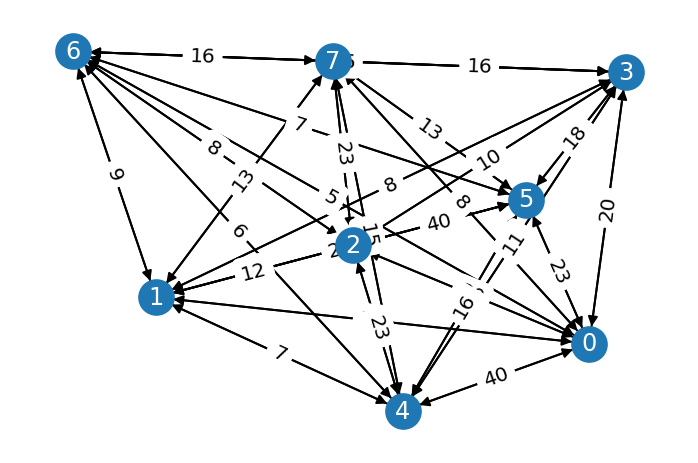}
\includegraphics[height=4cm]{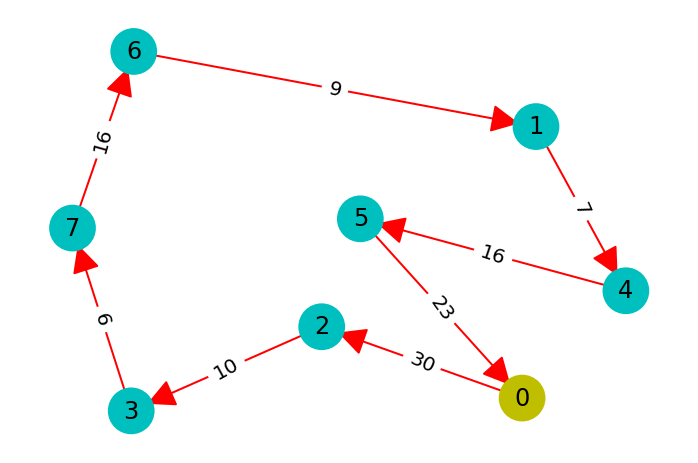}
\includegraphics[height=4cm]{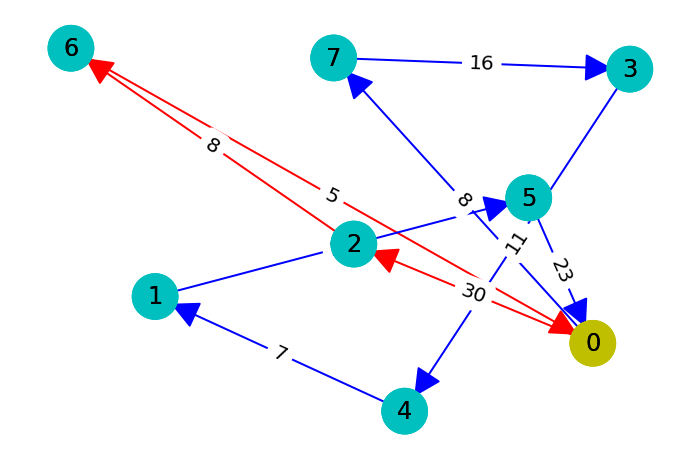}
\includegraphics[height=4cm]{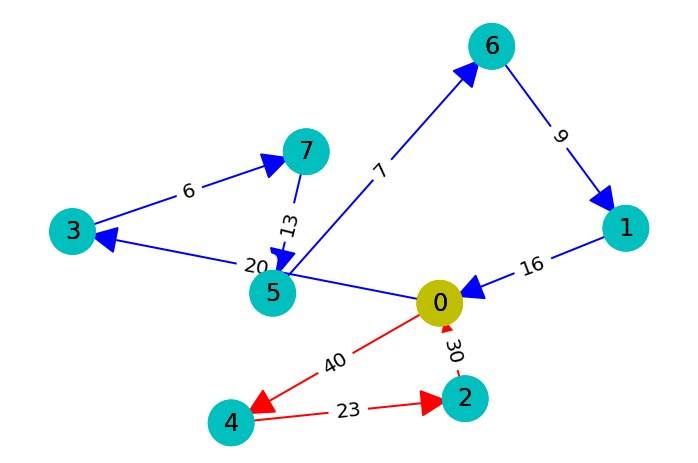}
\includegraphics[height=4cm]{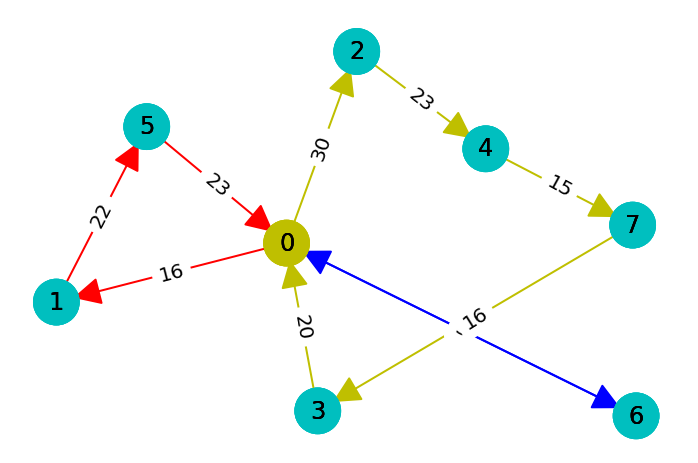}
\includegraphics[height=4cm]{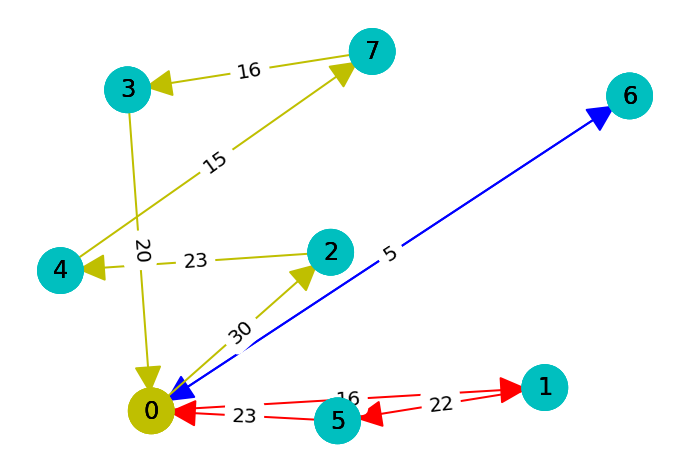}
\caption{In these graphs, we can observe the results of the algorithm in different scenarios. A different colour represents each qRobots; qRobot number $ 1 $ is red, next is blue, and the third is yellow, so on. While the depot is the $ 0 $ node in yellow color and the rest of the nodes are represented in blue. The weights of each item (not normalized) in kg are $ w_{0} = 0, w_{1} = 8, w_{2} = 8, w_{3} = 3  w_{4} = 3, w_{5} = 1, w_{6} = 2$ and $w_{7} = 4 $. The maximum capacity of each qRobots is $ 45 $.
 In this case, we have $ 7 $ items and the possibility of using up to $ 4 $ qRobots.
Reading the images from left to right, we see that the nodes and their respective distances are shown in the first image. In the second image, show the result of the algorithm having a qRobot. In the third and fourth images, we can see two different cases solved by two qRobots. And finally, in the fifth and sixth images, we can see two other issues solved by three qRobots.
It is important to highlight that our algorithm in this proof of concept minimizes the distance travelled and optimizes the number of qRobots necessary to solve the cases presented. If it judges that the task can be performed with a single qRobot, it will not send $ 2 $ qRobots.}
\label{fig:Results_qRobots_7}
\end{figure*}

\begin{figure}[!ht]
\centering
\includegraphics[width=0.45\textwidth]{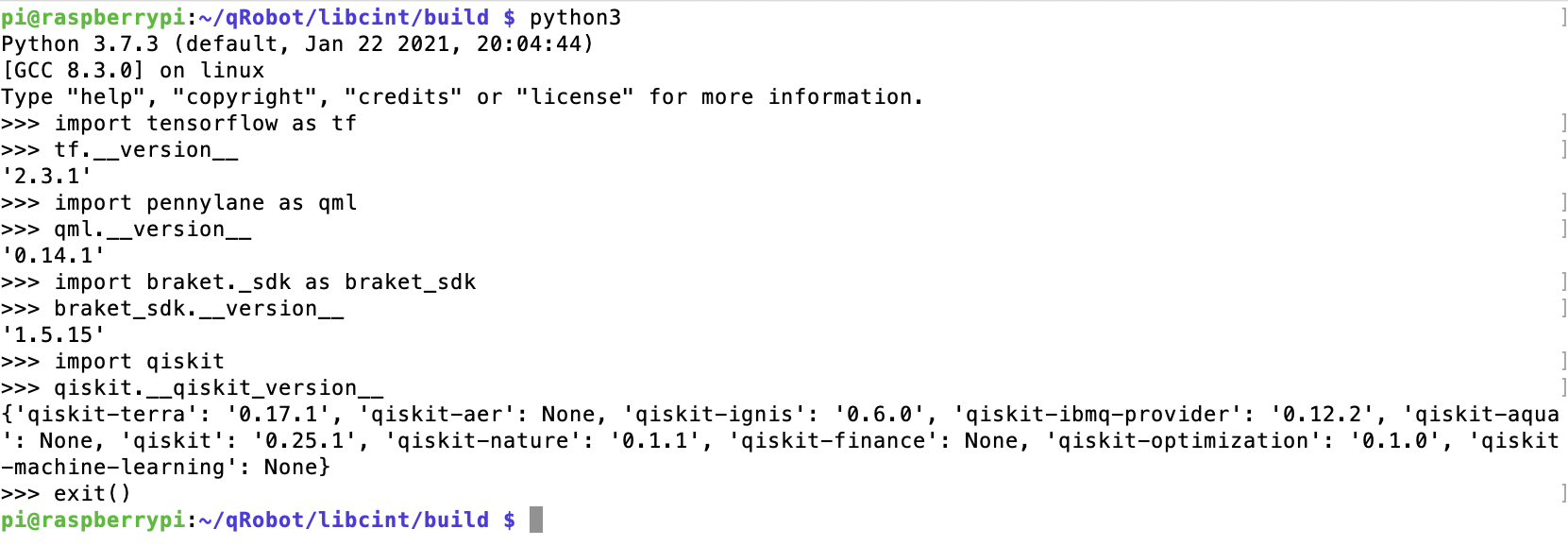}
\caption{This figure shows that we judge important environments to carry out quantum computing to robotics and beyond. We can see the correct installation of TensorFlow 3.2.1 as required for all gradient operations; see the installation of Pennylane version 14.1, the installation of the latest version of Amazon Braket, and all the packages of the newest version of qiskit 0.25 minus the qiskit-machine-learning package.}
\label{fig:installPackagesTerminal}
\end{figure}

\begin{figure}[!ht]
\centering
\includegraphics[width=0.45\textwidth]{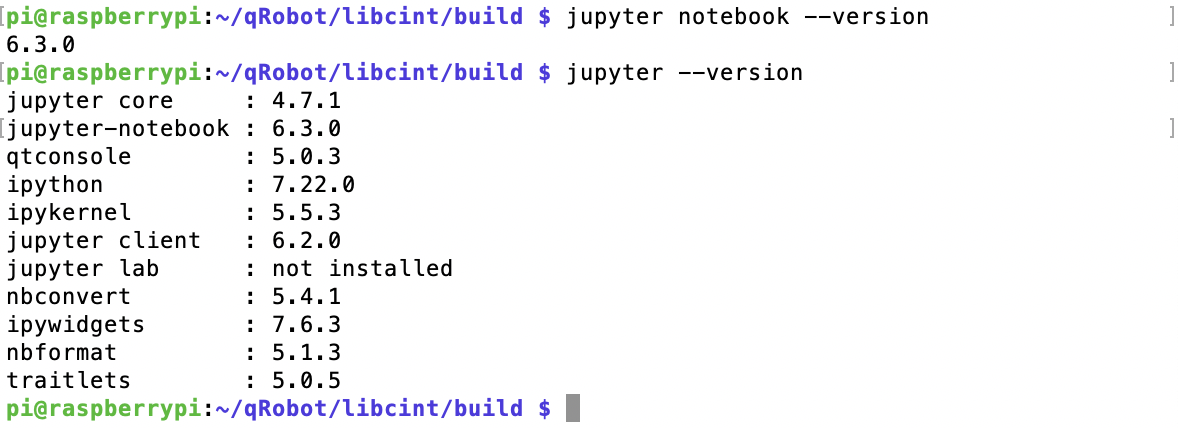}
\caption{In this figure, we can see the correct installation of the Jupyter package and the Jupyter notebook that has been our environment of proof of concept. With this, everything is ready to import or write code in the different frameworks mentioned above (IMBQ, AWS-Braket, Pennylane and D-Wave).}
\label{fig:Jupyter_Package}
\end{figure}

\begin{figure}[!ht]
\centering
\includegraphics[width=0.45\textwidth]{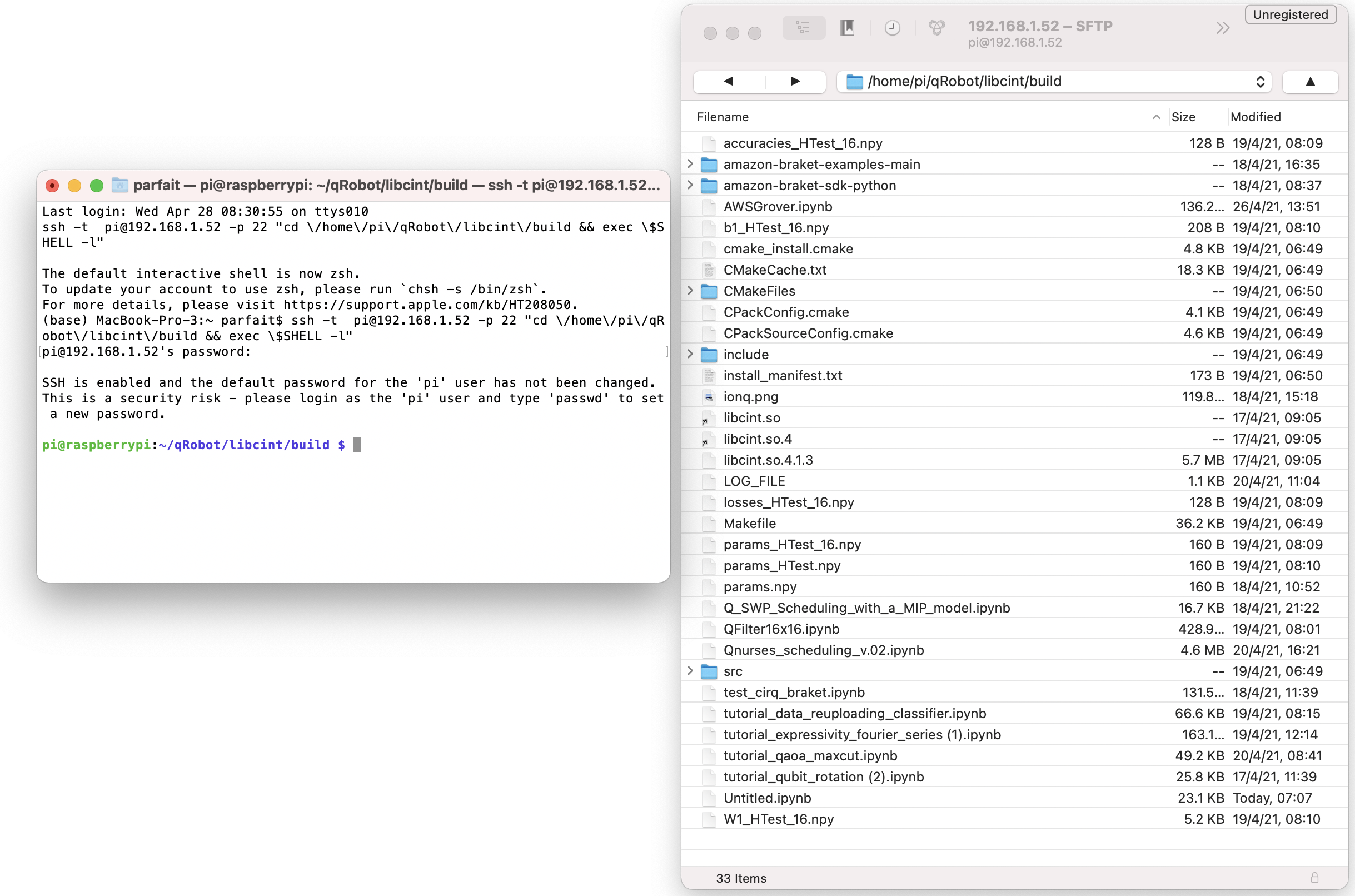}
\caption{This figure shows the files window through the CyberDuck client SSH \cite{T_cyberduck} viewer with the directory and file structure. And on the left, you can see the terminal that gives access to the qRobot. To access the qRobot by SSH, the username and password are required. Everything is configurable \cite{config_Rasp}.}
\label{fig:AccessSSH}
\end{figure}

\begin{figure*}[!ht]
\centering
\includegraphics[width=0.45\textwidth]{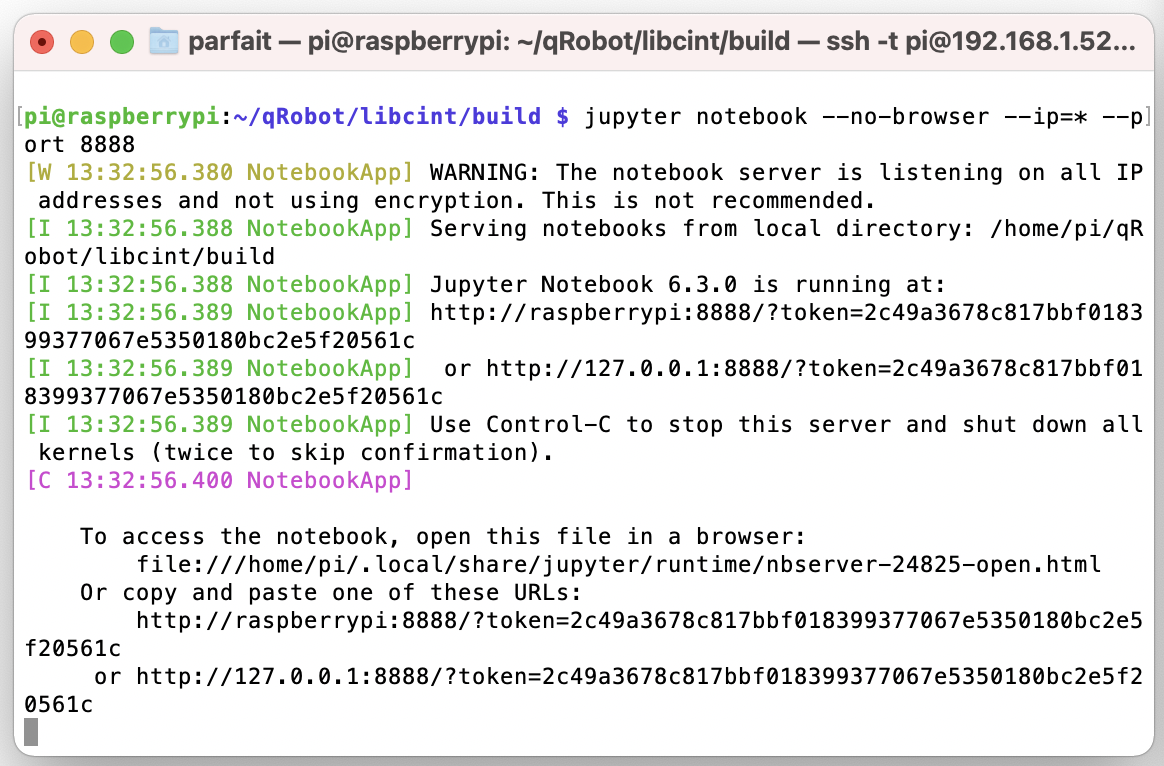}
\includegraphics[width=0.45\textwidth]{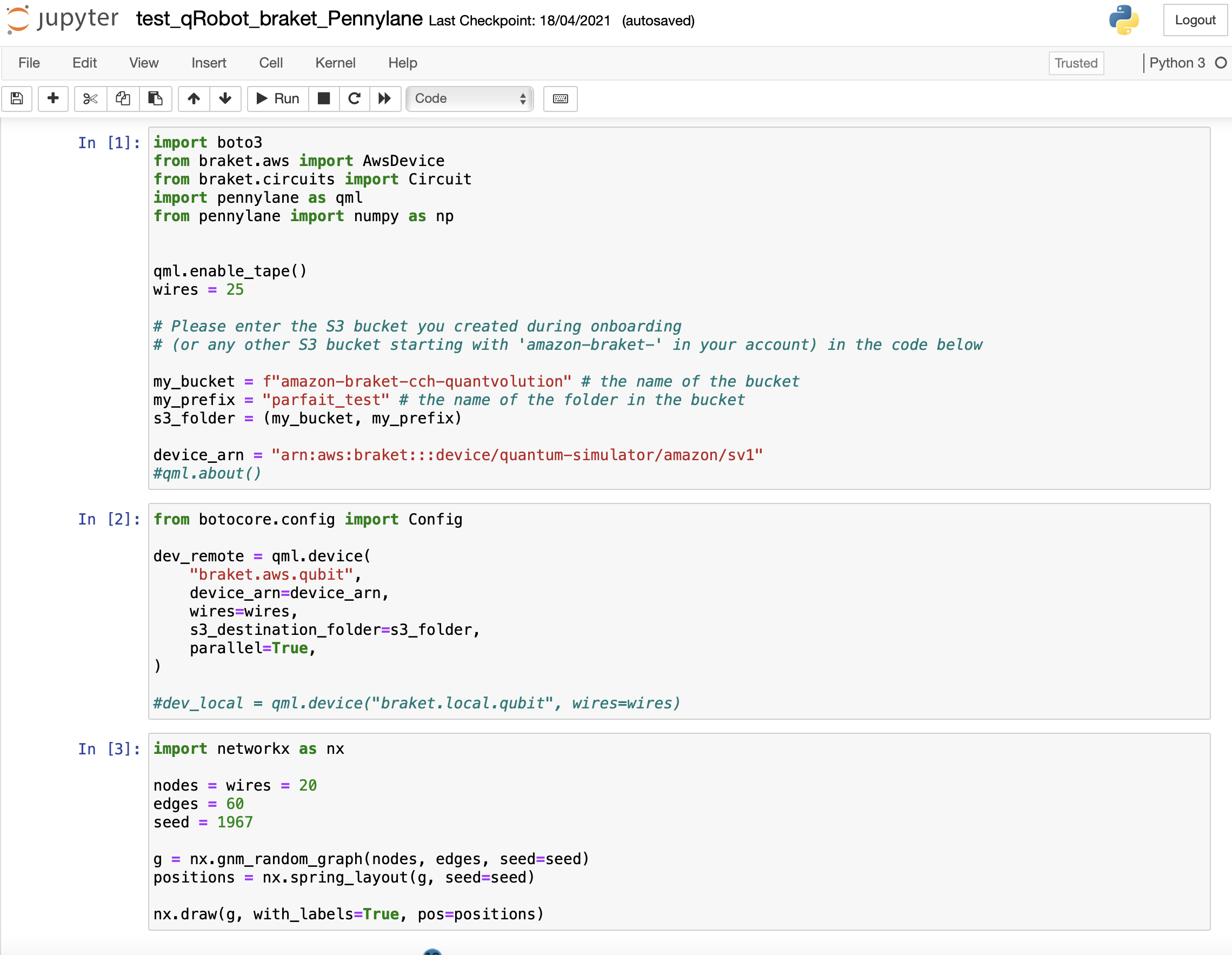}
\includegraphics[width=0.45\textwidth]{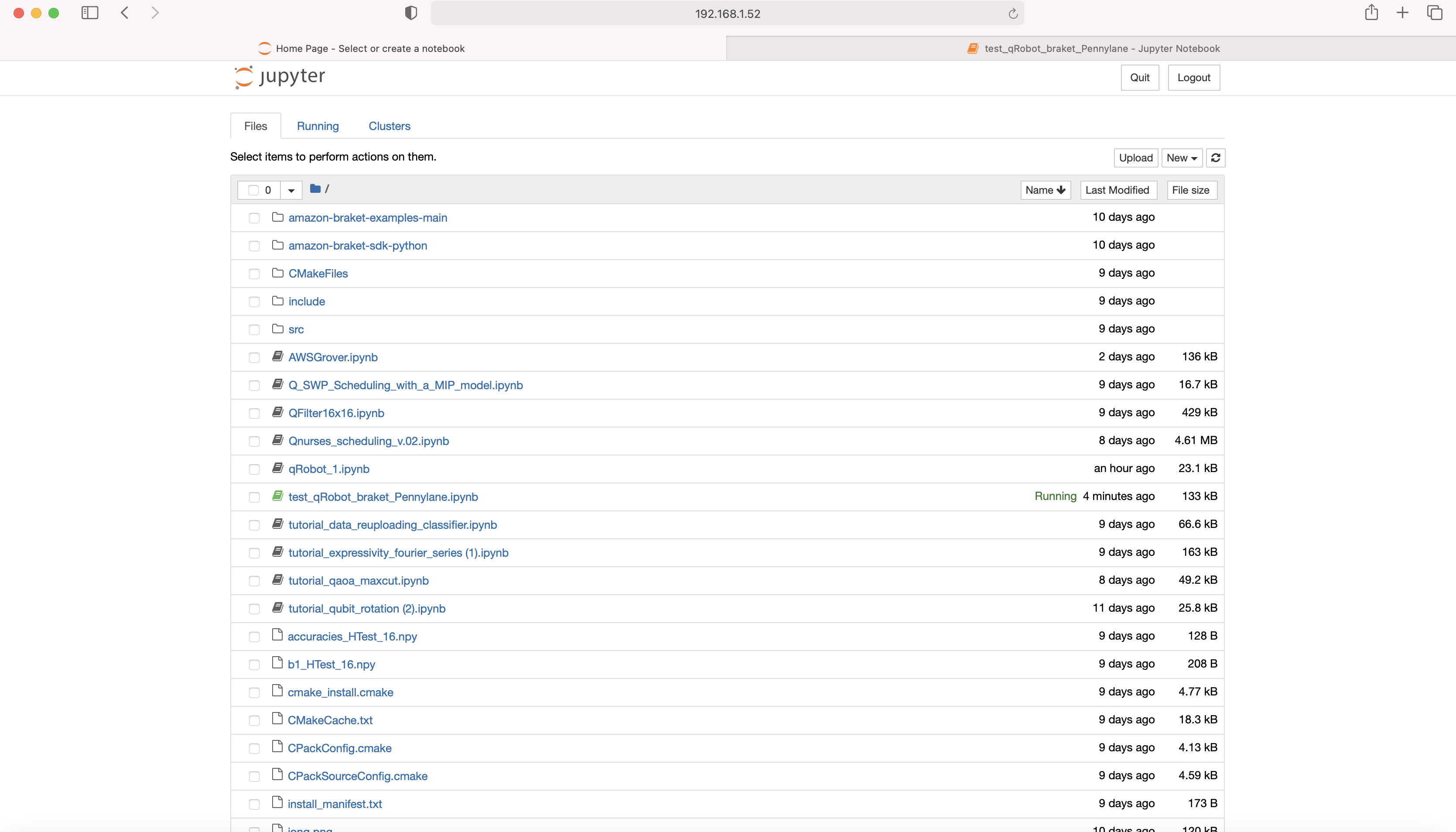}
\includegraphics[width=0.45\textwidth]{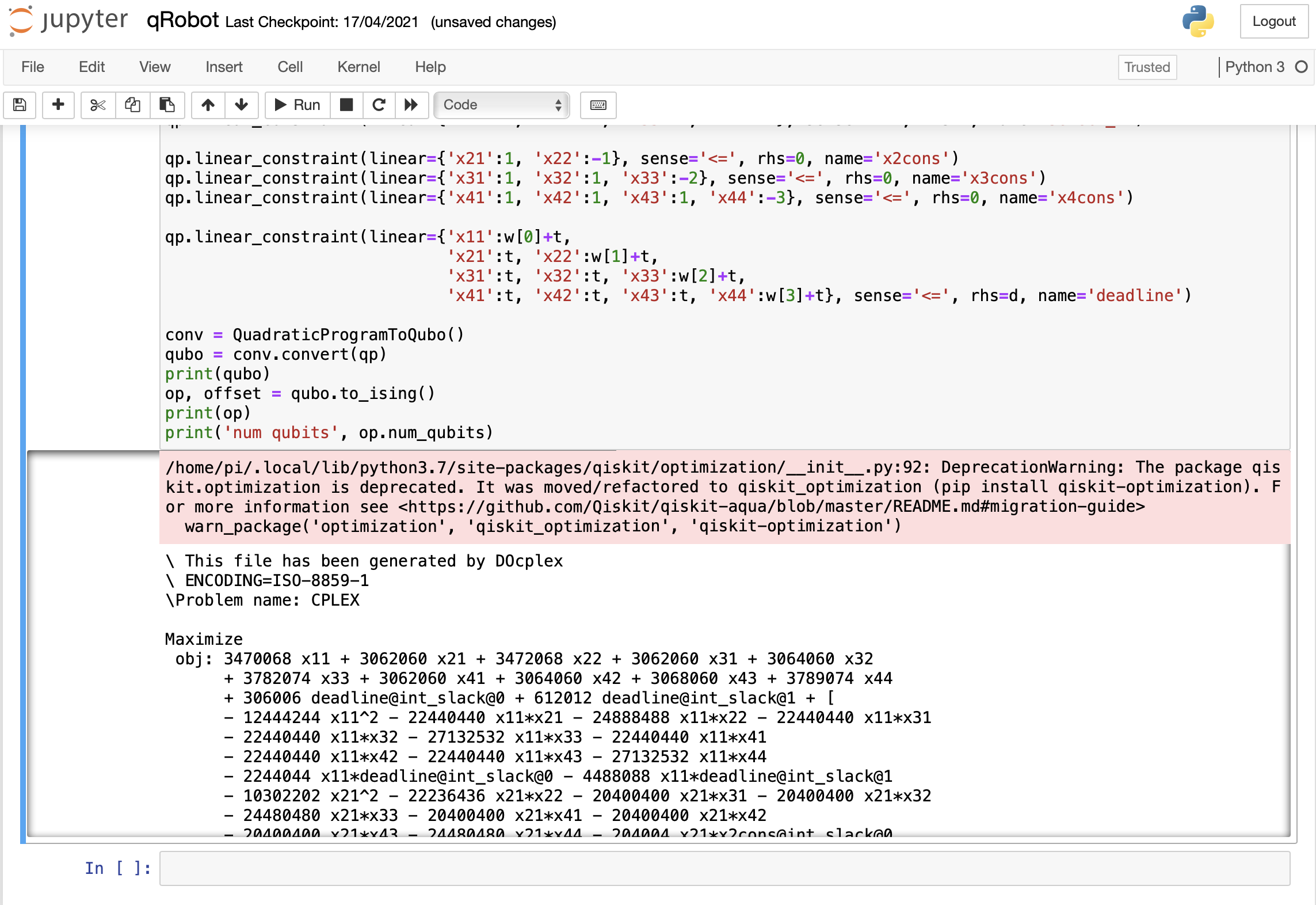}
\caption{In these figures, we see several notebook tests in operation. Works with the Qiskit, Pennylane\cite{bergholm2020pennylane} and AWS-Braket frameworks\cite{AWS_Braket}. It was also tested with quantum computers, Rigetti\cite{sete2016functional}, qiskit\cite{Qis21,mckay2018qiskit} and D-Wave\cite{dwave_computer}. In the figure of the terminal, you can observe the executions in progress. We can see from qiskit the docplex\cite{docplex} in execution. From AWS and Pennylane \cite{PennyL-AWS_Braket,SV1_AWS_Braket} we can see how to call the quantum device from the Raspberry.}
\label{fig:JupyterWorking}
\end{figure*}

Before analyzing in detail all the results of our proof of concept, it is of the utmost importance that we validate its operation globally and affirm that qRobot does meet our expectations and works as we expected.
Let's split the results of this proof of concept in two. 1, the configuration and conversion results of the Raspberry Pi 4 in a quantum computing environment (Fig.\eqref{fig:installPackagesTerminal} to Fig.\eqref{fig:JupyterWorking}) and 2, the picking and batching algorithm results represented by tables \eqref{tab:Benchmark_qrobot_simulators} to \eqref{tab:Table_benchmark_qRobot_With_La} on one side and Fig.\eqref{fig:Results_qRobots_4} and \eqref{fig:Results_qRobots_7} on the other.

The steps to convert the Raspberry Pi 4 into a "quantum computer" are in the Appendix \ref{sec:InstallationARM64}. 

Table \eqref{tab:Benchmark_qrobot_simulators} shows the experimentation results by setting the number of qRobots as their capacities (maximum load) at $1$ and $45$, respectively. We compare the execution time of our algorithm with different public access simulators on the market during this experimentation, solving the problem of picking and batching. We observed that, for issues of this nature, and especially due to the number of qubits required in each scenario, the behaviour of the D-Wave is the desired one at the temporal level, comparing it with Gate based Quantum Computing. However, it should be taken into account that, for experiments with numbers of qubits less than $20$, the behaviour of these simulators is equated with the D-Wave. This experimentation helps to have a clear vision about the feasibility of this proof of concept.

Continuing with the analysis of the results, table \eqref{tab:Table_benchmark_qRobot} shows us the computational results of our picking and batching algorithm considering $ 1 $ qRobot through AWS-Braket and on the real quantum computer D-Wave Advantage\_system1\cite{Zaborniak_2021}. The time value is an average and not counting latency time, job creation, and job return time. 

We also analyze the latency time when running the algorithm from the qRobot to the quantum computer. The quantum computer was in Oregon (US) and our qRobot in Barcelona (Spain) in the tests we've done. Out of all the tests we've run, we've had an average latency time of around $ 2 $ seconds plus all job management processes rising to roughly $ 8$ seconds. For the number of qubits greater than $30$, it is very convenient to use AWS-Braket (Advantage\_system1.1) instead of Qiskit or Pennylane for the number of qubits and the execution time; it is differentially better. This scenario makes very viable the use of quantum in robotics. For tests with a value of $ M $ less than those in the table, the number of qubits is relaxed, and the execution time is improved. This leads us to normalize the weights of the batches. Since the number of qubits follows the formula $ K (n + 1) (n + 2) + K \lceil log_ {2} M \rceil $, where the $K \lceil log_ {2} M \rceil$ qubits are needed as ancillaries qubits.

We also analyze the quantum real-time execution deeply through table \eqref{tab:Table_benchmark_qRobot_With_La}. We have measured the execution time without counting the latency time, creating jobs, and returning the work.

Fig.\eqref{fig:Results_qRobots_4} offers us the algorithm results in different scenarios where we analyze some important case, which helped us determine viable strategies within our proof of concept.  It is important to note that our algorithm minimizes the distance travelled and optimizes the number of qRobots. The Fig.\eqref{fig:Results_qRobots_7} repeats almost the same scenario but now considering $ 7 $ items with the same number of qRobots.

\begin{table*}[t!]
\centering
\begin{tabular}{ |c|c|c|c|c|  }
 \hline
 \multicolumn{5}{|c|}{The benchmark of the qRobot’s algorithm in different quantum simulators.} \\
 \hline
 \# of items & \# qubits & DWave - Time(s)  & Ibmq\_qasm\_simulator - Time(s) & Pennylane - Time(s) \\
 \hline
 2   & 18 & 1.92  &  1.89  & 1.94 \\
 3   & 26 & 3.2  &  737.46  & 656.93 \\
 4   & 36 & 4.88 &  -  & - \\
 5   & 48 & 7.60 &  -  & - \\
 6   & 62 & 11.16 &  -  & -  \\
 7   & 78 & 15.89 &  - & - \\
 8   & 96 & 21.72 &  -  & -\\
 9   & 116 & 30.18 &  - & -\\
 10  & 138 & 43.29 &  - & - \\
 11  & 162 & 53.28 &  - & - \\
 12  & 188 & 63.45 &  - & - \\
 \hline
\end{tabular}
\caption{ In this experimentation, both the number of qRobots and their capacities (maximum load) are fixed and are worth $1$ and $45$ respectively. We compare the execution time of our algorithm in the different public access simulators in the market, solving the picking and batching problem. We see that for issues of this nature, and especially for the number of qubits required in each scenario, the behaviour of the D-Wave is the desired one at the temporal level, comparing it with technologies based on quantum gates. However, it should be noted that for the experiments on numbers of qubits less than $20$, the behaviour of these simulators is equated with the D-Wave.
This experimentation helps to have a clear vision about the feasibility of this proof of concept.}
\label{tab:Benchmark_qrobot_simulators}
\end{table*}

\begin{table*}[!t]
    \centering
    \begin{tabular}{c|c|c|c|c|c|c}
         &  & & \textbf{ AWS-Braket\cite{AWS_Braket} } & \textbf{ ibmq\_qasm\_simulator\cite{Qis21} } & \textbf{ Pennylane\cite{bergholm2020pennylane} } \\
        \textbf{\#\ of items} & \textbf{qRobot's Capacity} & \textbf{\#\ qubits} & \textbf{ Average Time (s) } & \textbf{ Average Time (s)} & \textbf{ Average Time (s) } \\
         \hline
        2  & 15 & 10 & $11.23$ & $0.053$ & $0.041$ \\
        3  & 15 & 16 & $22.96$ & $0.40$ & $0.27$ \\
        4  & 15 & 24 &$33.07$ & $537.46$ & $480$ \\
        5  & 15 & 34 &$57.93$ & $-$ & $-$ \\
        6  & 15 & 46 &$118.41$  & $-$  & $-$ \\
        7  & 15 & 60 &$145.83$ & $-$ & $-$ \\
        8  & 15 & 76 &$296.81$ & $-$ & $-$ \\
        9  & 15 & 94 &$335.64$& $-$ & $-$ \\
        10 & 25 & 115 & $427.36$ & $-$ & $-$ \\
        11 & 25 & 137 & $650.25$ & $-$ & $-$ \\
        12 & 25 & 161 & $908.71$ & $-$ & $-$ \\
    \end{tabular}
    \caption{ Table of the computational results of our picking and batching algorithm on only $1$ qRobot. The value of time is an average and includes the waiting time, queue, execution and return of the solution. In the case of K is equal to $2$ for $9$ items with the qRobot capacity equal to $15$, the number of qubits is $188$. The execution time is on AWS Braket and on the D-Wave Advantage\_system1 quantum computer. We can realize that there is a latency time in executing the algorithm from the qRobot to the real quantum computer. In the tests we've done, the quantum computer is in the US West (Oregon). Of all the tests that we have done, we have had an average latency time of about $2$ plus all the work management processes that rises more or less to about $8$ seconds. For the number of qubits exceeding 30, it is very convenient to use AWS-Braket (Advantage\_system1.1)\cite{Zaborniak_2021} instead of Qiskit or Pennylane. By the number of qubits and the execution time, that is differentially better. This scenario makes the use of quantum in robotics very viable. For the tests with a value of $M$ lower than those in the table, the number of qubits is relaxed, and the execution time is improved. This leads us to normalize the weights of the batches. Since the number of qubits follows the formula $K(n+1)(n+2) + K \lceil log_{2}M\rceil$.}
    \label{tab:Table_benchmark_qRobot}
\end{table*}

\begin{table*}[!t]
    \centering
    \begin{tabular}{c|c|c|c|c|c|c}
         &  & & \textbf{ AWS-Braket\cite{AWS_Braket} } & \textbf{ ibmq\_qasm\_simulator\cite{Qis21} } & \textbf{ Pennylane\cite{bergholm2020pennylane} } \\
        \textbf{\#\ of items} & \textbf{qRobot's Capacity} & \textbf{\#\ qubits} & \textbf{ Average Time (s) } & \textbf{ Average Time (s)} & \textbf{ Average Time (s) } \\
         \hline
        2  & 15 & 10 & $0.13$ & $0.053$ & $0.041$ \\
        3  & 15 & 16 & $0.31$ & $0.40$ & $0.27$ \\
        4  & 15 & 24 &$1.69$ & $537.46$ & $480$ \\
        5  & 15 & 34 &$7.93$ & $-$ & $-$ \\
        6  & 15 & 46 &$11.31$  & $-$  & $-$ \\
        7  & 15 & 60 &$22.30$ & $-$ & $-$ \\
        8  & 15 & 76 &$36.11$ & $-$ & $-$ \\
        9  & 15 & 94 &$54.01$& $-$ & $-$ \\
        10 & 25 & 115 & $80.40$ & $-$ & $-$ \\
        11 & 25 & 137 & $139.67$ & $-$ & $-$ \\
        12 & 25 & 161 & $195.60$ & $-$ & $-$ \\
    \end{tabular}
    \caption{ In this table, we only consider the running time of the quantum algorithm on the real quantum computer from the qRobot (Advantage\_system1.1\cite{Zaborniak_2021}), not counting latency time, job creation, and job return time.}
    \label{tab:Table_benchmark_qRobot_With_La}
\end{table*}

\section{Discussions}\label{sec:Discussions}
We have achieved that, given a warehouse with a single robot, a list of several products with their respective loads and a list of batches, our system minimizes the distance to collect all the products and prepare the batches. This formulation solves the order in which the robot could manage all the products and make the batches passing through the depot. 
Another important achievement that offers this approach is that each robot makes a single trip. However, it is possible to band the code so that if we find ourselves in a situation where there are many batches to create and few robots to do the picking, these robots can be made to make the necessary trips if we have $ k $ qRobots that make at most one trip (we will never need more with $n$ batches). In this way, we will obtain all the packages for trips that we are interested in doing. A more understandable way of explaining it would be that when the first qRobot has finished its journey, it should only be ordered to do the one that would have made the qRobot $ k + 1 $, which does not exist. And so on with all the qRobots $ k + 2 $, $ k + 3 $, $ k + 4 $ ... until all scheduled batches are finished.

Right now, in addition to the processor, quantum computing simulation is closely related to memory. What takes up memory is to simulate a quantum computer, but the quantum computer does not need that memory, so it is assumed that it will end up being better.
In this proof of concept, using 8GB of RAM on the Raspberry Pi 4, we got the following results. The algorithm of collection and generation of packages take between $2$ and $450$ seconds to generate the batches and picking. If you want the qRobot to do all these tasks, we need to calculate the path before forming the packs.
That said, we must bear in mind that if what we want is to recalculate new routes when the robot has already left, we must take into account a lower latency time but close to said interval. A possible solution would be to choose a Raspberry with more RAM capacity. For example, if we had a 64GB Raspberry Pi, this time would be cut to 2/8, and it would take approximately 56.25 seconds (less than a minute) to create the batches. However, in this era of quantum computing, it is not representative to compare times since the computational differences will be noticed when the problems begin to grow, not on the small scales that we are currently dealing with.

Effective viability for today's warehouses would consist of splitting the tasks of the robots and having a qRobot that centralizes all the requests and passes them to the fleet of $ n $ qRobots so that they collect the products belonging to each batch.

We also did tests and developed a system that allows us to model the problem and run it on a Dwave.
Despite the optimization of the algorithm, the number of necessary qubits ($K(n+1)(n+2) + K \lceil log_{2}M\rceil$) and the need for low latency make this code adapted to the Annealing model. For this reason, we have prepared the Raspberry PI so that it can run D'wave directly and under Amazon-braket-ocean-plugin. For more information, see the steps in Appendix A.
With this scenario, one could have a "reasonable" latency for low data volume. Things that today, computers based on quantum gates cannot offer.

\section{Conclusions and further work}\label{sec:Conclusions}

As we have seen, the problem raised throughout this work offers us an efficient way of managing a series of $K$ qRobots to collect a set of orders, optimizing the number of robots used. The provided approach applies to a “central computer” capable of carrying out all the calculations and then giving each of the robots’ orders. However, when we begin to deal with very large problems both in the number of products and in the number of robots, the number of qubits required will tend to grow too large. A possible solution is to distribute the calculation of a central computer to each of the robots in such a way that each one has to calculate its route given a list of products to be collected. In this case, the equations of the problem would not change, just take $ K = 1 $ for each qRobot and apply the technique mentioned at the beginning of the discussion. Although it may not be possible to reach such the best solutions, this process of distribution of the calculation would suppose a significant computational cost reduction despite the need to create the batches beforehand. This search for batch creation will be studied in future projects.
On the other hand, it is important to note that the problem dealt with has a QUBO-type formulation, which allows it to be executed in annealing-type quantum computers. This makes a big difference in today's era (NISQ) as we have managed to work with $ ~ $ 200 qubits versus the $ ~ $ 30 qubits that we would have with a gate-based quantum computer. Finally, note that the defined problem seeks to minimize the total distance travelled by the robots, making it worthwhile for not all the robots to come out. For future line, we will address the same problem. Still, we will try to reduce the total times instead of the distance travelled (as done in this previous work\cite{Atc202}) since this situation is also very important in warehouse logistics.

\acknowledgments
The authors greatly thank the AWS-Braket and IBM team, mainly Simone Severini and Steve Wood respectively. P.A. thanks Jennifer Ramírez Molino for his support and comments on the manuscript.
\\

\textbf{Compliance with Ethics Guidelines}\\
\\
Funding: This study was not funded by any grant. Conflict of interest: P. Atchade-Adelomou, G. Alonso-Linaje, J. Albo-Canals and D. Casado-Faulí, state that there are no conflicts of interest.
Ethical approval: This article does not contain any studies with human or animal subjects.
Informed consent: Informed consent was obtained from all individual participants included
in the study.

\appendix \label{sec:Appendix}

\section{Installation of ARM64 on Raspberry Pi 4}\label{sec:InstallationARM64}
This section will describe step by step and delve into how installing and running Pennylane, AWS-Braket, D-Wave-Ocean, Qiskit, on a Raspberry Pi 4 under the ARM64\cite{jiang2020power} operating system torn it into a quantum computing simulator and use it to access real quantum computers from IBMQ\cite{Qis21,mckay2018qiskit}, AWS-Braket\cite{AWS_Braket}, D-Wave\cite{dwave_computer}, and Regetti\cite{sete2016functional}. These frameworks and packages are required for the proof of concept that we propose.
\begin{figure}[!ht]
\centering
\includegraphics[width=0.45\textwidth]{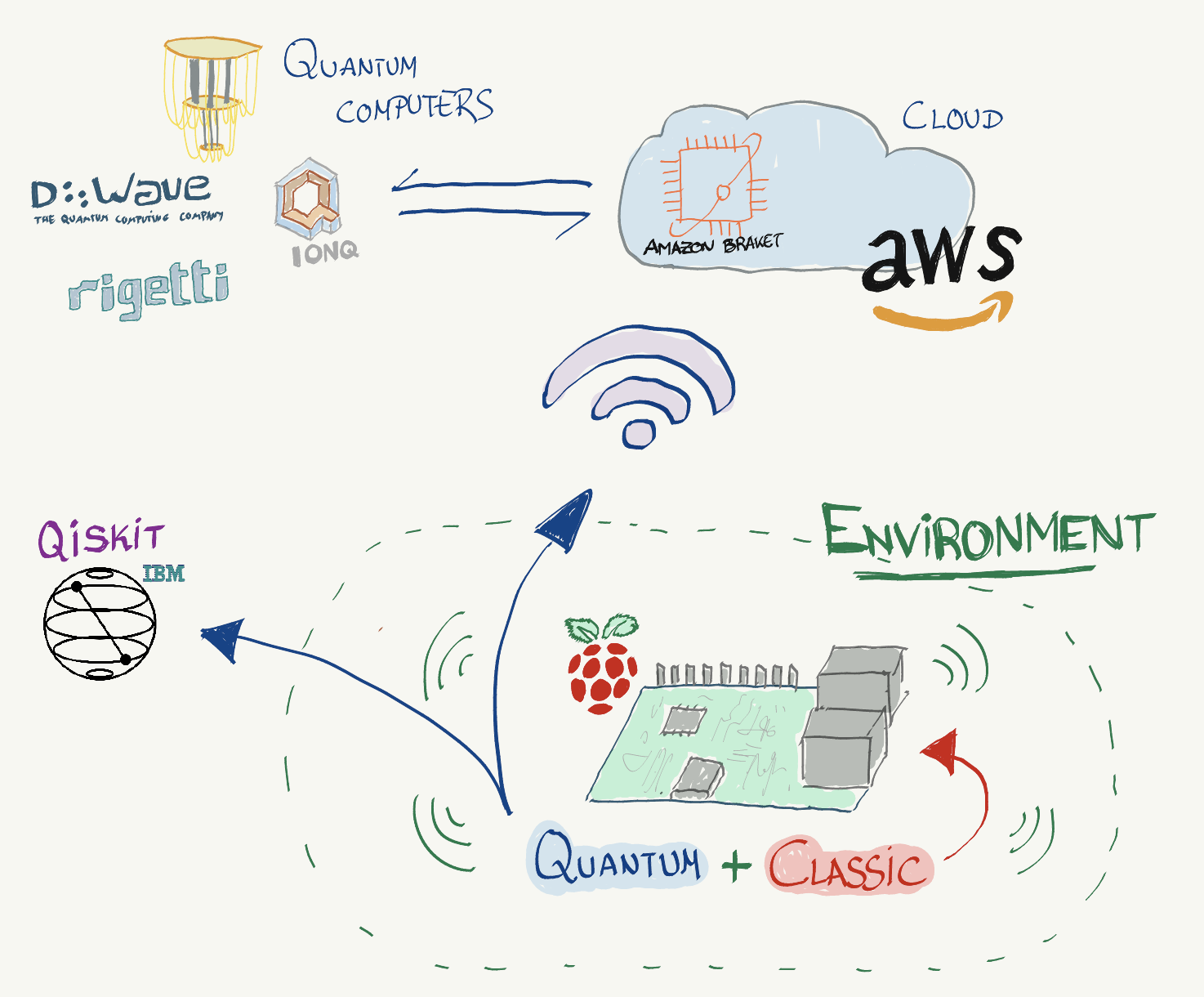}
\caption{We have installed the following frameworks successfully (Qiskit, Pennylane, AWS-Braket) on our Raspberry Pi 4 under the ARM64 operating system.}
\label{fig:qRobot_platform}
\end{figure}

\definecolor{codegreen}{rgb}{0,0.6,0}
\definecolor{codegray}{rgb}{0.5,0.5,0.5}
\definecolor{codepurple}{rgb}{0.58,0,0.82}
\definecolor{backcolour}{rgb}{0.95,0.95,0.92}

\lstdefinestyle{mystyle}{
  backgroundcolor=\color{backcolour},   commentstyle=\color{codegreen},
  keywordstyle=\color{magenta},
  numberstyle=\tiny\color{codegray},
  stringstyle=\color{codepurple},
  basicstyle=\ttfamily\footnotesize,
  breakatwhitespace=false,         
  breaklines=true,                 
  captionpos=b,                    
  keepspaces=true,                 
  numbers=left,                    
  numbersep=5pt,                  
  showspaces=false,                
  showstringspaces=false,
  showtabs=false,                  
  tabsize=2
}

\lstset{style=mystyle}
\begin{lstlisting}[language=Python, caption=Installation Raspberry Pi 4 ARM64]
Steps
1. Download the latest image of Raspberry Pi ARM64
2. Initial setup of a headless Raspberry Pi
3. Setup of the Python environment and TensorFlow 2.3.1
4. Manual installation of some dependencies
5. Installation of the Qiskit elements
6. Installation of the Pennylane elements
7. Installation of the Amazon elements
8. Setup of Jupyter Notebooks
9. Enable remote desktop access using VNC
10. Test Jupyter notebook codes
11. Install DWave framework and Amazon-braket-ocean-plugin

Note: The actual version of the ARM64 for Raspberry Pi 4 is not stable. https://www.raspberrypi.org/forums/viewtopic.php?t=275370 

1. Download the latest image of Raspberry Pi ARM64
Download the image from: https://downloads.raspberrypi.org/raspios_arm64/images/raspios_arm64-2021-04-09/ 

2. Initial Setup of a headless Raspberry Pi
We want to setup a headless Raspberry Pi (i.e. without display, keyboard, mouse), and also not use display/keyboard/mouse during the setup procedure.
Creating an SD card with the initial OS is described at https://www.raspberrypi.org/documentation/installation/sdxc_formatting.md
We will use the Raspberry Pi Imager and choose %Raspberry Pi OS Desktop (64-bit)% to write the image to the SD card. It is recommended to use the Desktop image vs. the other alternatives.

Prepare for wireless boot
According to https://www.raspberrypi.org/documentation/configuration/wireless/headless.md, we create a file wpa_supplicant.conf in the root directory of the SD card with the following content (replace (DE) with the appropriate country code, and (SSID)and (WLAN PASSWORD) with the SSID and password for our WLAN access point):
country=DE
ctrl_interface=DIR=/var/run/wpa_supplicant GROUP=netdev
update_config=1
network={
   ssid="SSID"
   psk="WLAN PASSWORD"
}
Now, we boot the Raspberry Pi, i.e., insert the SD card into the Raspberry Pi and connect it to a power supply.

3. Set up the Python environment
Do not use conda/anaconda/berryconda as recommended on other hardware platforms for Qiskit you can use the new virtual environment as you judge it suitable.

TensorFlow 2.3.1 for Python 3.
The whole shortcut procedure is found below. The wheel was too large to store at GitHub, so Google drive is used. Please make sure you have the latest pip3 and python3 version installed; otherwise, pip may come with the message ".whl is not a supported wheel on this platform".

Check your Python3 version. Each version needs a unique wheel. Currently, the Raspberry Pi 64-bit operating system uses Python 3.7.3. So you need to download Tensorflow-2.3.1-cp37-cp37m-linux_aarch64.whl. Undoubtedly, the Python version will upgrade over time and you will need a different wheel. See out GitHub page for all the wheels.
# get a fresh start (remember, the 64-bit OS is still under development)
$ sudo apt-get update
$ sudo apt-get upgrade
# install pip and pip3
$ sudo apt-get install python-pip python3-pip
# remove old versions, if not placed in a virtual environment (let pip search for them)
$ sudo pip uninstall tensorflow
$ sudo pip3 uninstall tensorflow
# install the dependencies (if not already onboard)
$ sudo apt-get install gfortran
$ sudo apt-get install libhdf5-dev libc-ares-dev libeigen3-dev
$ sudo apt-get install libatlas-base-dev libopenblas-dev libblas-dev
$ sudo apt-get install liblapack-dev
# upgrade setuptools 47.1.1 -> 50.3.0
$ sudo -H pip3 install --upgrade setuptools
$ sudo -H pip3 install pybind11
$ sudo -H pip3 install Cython==0.29.21
# install h5py with Cython version 0.29.21 (6 min @1950 MHz)
$ sudo -H pip3 install h5py==2.10.0
# install gdown to download from Google drive
$ pip3 install gdown
# copy binairy
$ sudo cp ~/.local/bin/gdown/usr/local/bin/gdown
# download the wheel
$ gdown https://drive.google.com/uc?id=1jbkp2rSZZ3YY-AM1vuHyB9hI05zrZGHg
# install TensorFlow (63 min @1950 MHz)
$ sudo -H pip3 install tensorflow-2.3.1-cp37-cp37m-linux_aarch64.whl

When the installation is successful, you should get the following screendump by executing: 
$ python3
>>> import tensorflow as tf
>>> tf.__version__
you may have 2.3.1

Now you may install the pyscf for more information: http://pyscf.org/pyscf/install.html#compiling-from-source-code

Prerequisites for manual install are
* 	CMake >= 3.10
* 	Python >= 3.6
* 	Numpy >= 1.13
* 	Scipy >= 0.19
* 	h5py >= 2.7
You can download the latest version of PySCF (or the development branch) from GitHub:
$ git clone https://github.com/pyscf/pyscf.git
$ cd pyscf
$ git checkout dev  # optional if you'd like to try out the development branch
Next, you need to build the C extensions in pyscf/lib:
$ cd pyscf/lib
$ mkdir build
$ cd build
$ cmake ..
$ make   #(30 min)

export PYTHONPATH=/opt/pyscf:$PYTHONPATH

please check if the package hs been installed successfully 
>>> import pyscf

Execute this:
cd pyscf/lib
sh _runme_to_fix_dylib_osx10.11.sh


4. Manual installation of some dependencies
 Based on and taking advantage of @Jan Lahmann, we need to install and configure some prerequisites first manually.
retworkx
We will install retworkx according to the instructions in https://retworkx.readthedocs.io/en/stable/README.html#installing-retworkx. First, install the rust language environment.
pi$ cd ~/qrobot
pip install setuptools-rust
curl -o get_rustup.sh -s https://sh.rustup.rs
sh ./get_rustup.sh -y
Now activate rust and install retworkx:
pi$ source ~/.cargo/env
pip3 install  retworkx


5. Installation of the Qiskit elements
After the pre-work we just completed, installing Qiskit should now be as simple as

pip3 install --force-reinstall pip
#pip3 install vaex
sudo apt install llvm-7-dev

#I recommend to install separely each paquet from qiskit. The version of the installed qiskit is 0.25.1
#In this version, you will not be able to install qiskit-machine-learning
pip3 install qiskit-aqua
pip3 install qiskit-aer
pip3 install 'qiskit[visualization]'

#Now, let us see what versions of Qiskit were installed:
pip3 list | grep qiskit
qiskit 0.25.1
qiskit-aer 0.8.1
qiskit-aqua 0.9.1
qiskit-finance 0.1.0
qiskit-ibmq-provider 0.12.2
qiskit-ignis 0.6.0
qiskit-nature 0.1.1
qiskit-optimization 0.1.0
qiskit-terra 0.17.1

python --version
>>>Python 3.7.3

Command "python setup.py egg_info" failed with error code 1 in /tmp/pip-install-eur2lck3/qiskit-aer/

6. Installation of the Pennylane elements
pip install pennylane --upgrade
pip install autograd

7. Installation of the Amazon elements
pip install amazon-braket-sdk
pip install amazon-braket-pennylane-plugin

Need to set if you specify directly with boto3, it would be like this but you are using PennyLane
https://boto3.amazonaws.com/v1/documentation/api/latest/guide/configuration.html
https://boto3.amazonaws.com/v1/documentation/api/latest/guide/configuration.html#using-a-configuration-file
aws_access_key_id and aws_secret_access_key will also be required, which are associated with AWS IAM User.

For that, you must need any ~/.aws/config file
Edit with:
cat ~/.aws/config

mkdir ~/.aws
touch ~/.aws/config
echo "[default]" >> ~/.aws/config
echo "region = us-east-1" >> ~/.aws/config
echo "aws_access_key_id = AKIAIOSFODNN7EXAMPLE" >> ~/.aws/config
echo "aws_secret_access_key = wJalrXUtnFEMI/K7MDENG/bPxRfiCYEXAMPLEKEY" >> ~/.aws/config

I use by default us-east-1, but the user can use the region he got.

the output format is like this example:
[default]
region = us-east-1
aws_access_key_id = AKIAIOSFODNN7EXAMPLE
aws_secret_access_key = wJalrXUtnFEMI/K7MDENG/bPxRfiCYEXAMPLEKEY

From here, you can be able to execute any code in Pennylane or AWS-Braket. For another platform like D-Wave, you may need to install:
pip3 install amazon-braket-ocean-plugin
pip3 install dwave_networkx
pip3 install minorminer
pip3 install dwave-ocean-sdk

8. Setup of Jupyter Notebooks 
pip3 install jupyter

Start Jupyter without a local browser and listen on port 8888 for a remote connect:
jupyter notebook --no-browser --ip=* --port 8888

Please follow the message from your prompt. It might be necessary to replace the hostname (raspberrypi) with the correct hostname in our local network or the IP address of the raspberry.

You can configure to access as local and access the Jupyter notebook interface using the URL http://raspberrypi:8888/ from a browser on our laptop. You will need to replace (raspberry) with the correct hostname or IP of the Raspberry Pi. For that, you need to execute the next command:

mkdir -p ~/qRobot/temp; cd ~/qRobot/temp;
jupyter notebook --no-browser

9. Enable remote desktop access using VNC
In addition to connecting to the Raspberry Pi via ssh, it might be useful to enable access with VNC to connect to a graphical desktop that is running locally on the Raspberry Pi. This is described at https://desertbot.io/blog/headless-raspberry-pi-4-remote-desktop-vnc-setup.
First, we enable VNC and change the screen resolution:
 sudo raspi-config
Select Interfacing Options
Select VNC
For the prompt to enable VNC, select Yes (Y)
For the confirmation, select Ok
Select Advanced Options
Select Resolution
Select anything but the default (example: 1280x720)
Select Ok
Select Finish, Yes to reboot

For my local ssh viewer, I used Cyberduck https://cyberduck.io/, but you can also use https://www.realvnc.com/en/connect/download/viewer/ and connect to the Raspberry Pi (enter the IP address in VNC viewer; enter login information). After the first connect, we will be asked to adjust some configurations (location settings, display settings, system update, etc.).

10. Test Jupyter notebook codes,

11.	Install DWave framework and Amazon-braket-ocean-plugin
As DWave, unfortunately, does not provide ARM wheels yet. That means that you need to build from the source distributions. Simultaneously, dimod requires boost (https://github.com/dwavesystems/dimod#installation), though they are planning to remove that dependency soon (https://github.com/dwavesystems/dimod/issues/618, https://github.com/dwavesystems/dimod/pull/748).
You can try installing boost (https://www.boost.org/) and then trying to install dimod again.
The simple way is by installing as apt-get install libboost-dev. So follow the next steps below. During these steps, you may need to upgrade your pip or NumPy.
apt-get install libboost-dev
pip3 install amazon-braket-ocean-plugin

After these steps, you must need to install the package from Dwave.

By installing dwave-tabu from source on master (we switched from swig to cython, but haven't released 0.4 yet):
pip install -U pip setuptools
USE_CYTHON=1 pip install -e git+https://github.com/dwavesystems/dwave-tabu.git#egg=dwave-tabu

after this, install pip3 install dwave-system
Then you already have your system ready to use DWave from your Raspberry PI 4.

\end{lstlisting}

\newpage
\bibliographystyle{unsrturl}
\bibliography{main}
\end{document}